%% file: higgs-pairs.tex
\documentclass[nofootinbib,prd,aps,superscriptaddress,preprintnumbers]{revtex4-1}
\pdfoutput=1
\usepackage[utf8]{inputenc}
\usepackage{amsmath,amssymb,euscript}
\usepackage{slashed}
\usepackage{color}
\usepackage{accents}
\usepackage{hyperref}
\usepackage{ulem}
\usepackage{epsfig}
\usepackage{varioref}
\usepackage{xcolor}
\usepackage{verbatim}

\def\beq{\begin{equation}}
\def\eeq{\end{equation}}

\newcommand{\tev}{\,\, \mathrm{TeV}}
\newcommand{\gev}{\,\, \mathrm{GeV}}

\newcommand{\met}{\slashed{E}_T}

\renewcommand{\emph}{\textit}

\graphicspath{{figures/}}

\begin{document}

\begin{flushright}

\end{flushright}

\title{Spotting hidden sectors with Higgs binoculars}
\hfill{TTP19-005}

\hfill{P3H-19-006}

\author{Monika Blanke}
\affiliation{Institute for Nuclear Physics (IKP), Karlsruhe Institute of Technology, Hermann-von-Helmholtz-Platz 1, D-76344 Eggenstein-Leopoldshafen, Germany}
\affiliation{Institute for Theoretical Particle Physics (TTP), Karlsruhe Institute of Technology, Engesserstra\ss e 7, D-76128 Karlsruhe, Germany}
\author{Simon Kast}
\affiliation{Institute for Theoretical Particle Physics (TTP), Karlsruhe Institute of Technology, Engesserstra\ss e 7, D-76128 Karlsruhe, Germany}
\affiliation{Institute for Theoretical Physics (ITP), Heidelberg University, Philosophenweg 12, D-69120 Heidelberg, Germany}
\author{Jennifer M. Thompson}
\affiliation{Institute for Theoretical Physics (ITP), Heidelberg University, Philosophenweg 12, D-69120 Heidelberg, Germany}
\author{Susanne Westhoff}
\affiliation{Institute for Theoretical Physics (ITP), Heidelberg University, Philosophenweg 12, D-69120 Heidelberg, Germany}
\author{Jos\'e Zurita}
\affiliation{Institute for Nuclear Physics (IKP), Karlsruhe Institute of Technology,
Hermann-von-Helmholtz-Platz 1, D-76344 Eggenstein-Leopoldshafen, Germany}
\affiliation{Institute for Theoretical Particle Physics (TTP), Karlsruhe Institute of Technology, Engesserstra\ss e 7, D-76128 Karlsruhe, Germany}

\vspace{1.0cm}
\begin{abstract}
\vspace{0.2cm}\noindent 
We explore signals of new physics with two Higgs bosons and large missing transverse energy at the LHC. Such a signature is characteristic of models for dark matter or other secluded particles that couple to the standard model through an extended scalar sector. Our goal is to provide search strategies and an interpretation framework for this new signature that are applicable to a large class of models. To this end, we define simplified models of hidden sectors leading to two different event topologies: \emph{symmetric decay}, i.e., pair-produced mediators decaying each into a Higgs plus invisible final state; and \emph{di-Higgs resonance}, i.e., resonant Higgs-pair production recoiling against a pair of invisible particles. For both scenarios, we optimize the discovery potential by performing a multi-variate analysis of final states with four bottom quarks and missing energy, employing state-of-the-art machine learning algorithms for signal-background discrimination. We determine the parameter space that the LHC can test in both scenarios, thus facilitating an interpretation of our results in terms of complete models. Di-Higgs production with missing energy is competitive with other missing energy searches and thus provides a new opportunity to find hidden particles at the LHC.
\end{abstract}
\maketitle


\section{Introduction}\label{sec:introduction}
\input{introduction}

\section{Simplified models of a hidden scalar sector}\label{sec:models}
\input{models}

\section{Higgs-pair production with missing energy at the LHC}\label{sec:signature}
 \input{signature}

\section{Multi-variate analysis and results}\label{sec:mva}
\input{mva}

\section{Dark matter}\label{sec:dm}
\input{dm}

\section{Conclusions}\label{sec:conclusions}
\input{conclusions}

\section{Acknowledgements}
\noindent We thank Martin Bauer, Ulrich Haisch, Jos\'e Miguel No and Tilman Plehn for helpful discussions. This project has been realized within the Heidelberg Karlsruhe Strategic Partnership HEiKA. We acknowledge HEiKA funding through the research bridge ``Particle Physics, Astroparticle Physics and Cosmology (PAC)'' that facilitated our collaboration.  The research of MB and SW is supported by the DFG Collaborative Research Center TRR 257 ``Particle Physics Phenomenology after the Higgs Discovery''.
SK acknowledges  the  support  of the DFG-funded Doctoral School ``Karlsruhe School of Elementary and Astroparticle Physics: Science and Technology'' (KSETA). SW acknowledges funding by the Carl Zeiss foundation through an endowed junior professorship (\emph{Junior-Stiftungsprofessur}).

\appendix

\section{UV completing the $Sgg$ coupling}
\label{app:UV}
\input{UV-completion}

\end{document}

%% file: introduction.tex
\noindent Postulating a hidden sector that interacts primarily with the Higgs boson is tempting for good reasons. Higgs couplings to new scalar standard-model (SM) singlets are renormalizable and secluded from visible matter~\cite{Patt:2006fw,OConnell:2006rsp}. An extended scalar sector can thus serve as a portal to a hidden sector~\cite{LopezHonorez:2006gr,MarchRussell:2008yu,Englert:2011yb,Athron:2017kgt}. At the LHC, the Higgs interaction with a hidden sector is best probed in signatures with one or two Higgs bosons. Searches for invisibly decaying Higgs bosons or for mono-Higgs production in association with missing transverse energy $\met$ are well-established parts of the LHC program. Invisible Higgs decays probe hidden sectors with particles significantly lighter than the Higgs boson. Mono-Higgs signals are often predicted in models that can also be probed in other channels, such as mono-jet production, mono-$Z$ production, or signatures with missing energy and several leptons and/or jets. For a review of missing energy searches at the LHC, we refer the reader to Refs.~\cite{Kahlhoefer:2017dnp,Abe:2018bpo} and references therein.

A signal of two Higgs bosons and missing energy is naturally predicted in the context of supersymmetry (SUSY), for instance from Goldstino production in models with gauge mediated SUSY breaking~\cite{Giudice:1998bp,Matchev:1999ft}, or from chain decays of superpartners into Higgs bosons and neutralinos in the minimal supersymmetric standard model (MSSM) and its extensions~\cite{Kang:2015nga,Bernreuther:2018nat,Titterton:2018pba}. More generally, di-Higgs plus missing energy is a signature of models with extra scalars~\cite{Arganda:2017wjh}, such as a pseudo-scalar portal to a dark sector~\cite{No:2015xqa}, axion-like particles~\cite{Brivio:2017ije}, massive right-handed neutrinos~\cite{Kang:2015nga,Kang:2015uoc} or in the framework of Little Higgs scenarios~\cite{Kang:2015nga,Chen:2018dyq}. Experimental searches for di-Higgs plus $\met$ production at the LHC have been performed for a signal of four bottom quarks and missing energy in the context of SUSY~\cite{Sirunyan:2017obz,Aaboud:2018htj}. This search targets a signature of Higgsino pair production, followed by a decay chain with Higgs bosons and Goldstinos in the final state~\cite{Matchev:1999ft}. Since the analysis is optimized for very light Goldstinos produced via this specific decay chain, its reinterpretation in other scenarios is limited. A systematic exploration of the di-Higgs plus $\met$ channel at the LHC is still lacking.
 
Our goal is to provide a minimal, simple framework to exploit the full potential of the LHC to search for new hidden sectors with a di-Higgs plus $\met$ signature. As a matter of fact, the search strategy for this signature strongly depends on the masses and decays of the relevant particles. Based on two main decay topologies, we define simplified models for $pp\to hh\chi\chi$ production, where $h$ is the SM Higgs boson and $\chi$ is invisible and stable at detector scales. Each model involves two scalar mediators $B$ and $A$, where $B$ couples to gluons and is heavier than two $A$ scalars. The first model, referred to as \emph{symmetric topology}, is inspired by electroweakino production in the MSSM. A pair of on-shell scalars $A$ is produced from the decay of $B$. Each $A$ subsequently decays into a Higgs boson and an invisible scalar $\chi$. The di-Higgs signature is thus generated by
\begin{align}
pp \to B \to AA \to (h\chi)(h\chi).
\end{align}
In the second model, referred to as \emph{resonant topology}, each of the pair-produced scalars $A$ decays into either two Higgs bosons or invisibly. The corresponding production chain is
\begin{align}
pp\to B \to AA \to (hh)(\chi\chi).
\end{align}
Such a topology is typical in scalar portal models. Since the definition of the two simplified models is based solely on the kinematic properties of the final state, LHC searches for these simple topologies can easily be recasted in terms of concrete models.

Our analysis focuses on the Higgs decay into bottom quarks, $h\to b\bar{b}$, which maximizes the event rates. The signal thus consists of four $b$-jets and a large amount of missing transverse energy. To reconstruct the two Higgs bosons from the four $b$-jets, we will make ample use of the mature analysis techniques for di-Higgs searches without associated missing energy. Due to its sensitivity to the Higgs self-interaction, Higgs pair production in the SM (see Refs.~\cite{ATLAS:2018otd,CMS:2018rig} for the latest experimental prospects) and beyond (for a review see e.g. Ref.~\cite{Zurita:2017sfg} and references therein) is a key target of the LHC program and proposed future colliders~\cite{Contino:2016spe}. The prospects to observe a signal of Higgs pairs has evolved from ``seemingly impossible"~\cite{Baur:2003gp} to a detailed investigation of the final states $b \bar{b}\,\tau^+ \tau^-$~\cite{Dolan:2012rv}, $b \bar{b}\,W W^*$\cite{Papaefstathiou:2012qe} and $b \bar{b}\,b \bar{b}$~\cite{deLima:2014dta}. This tremendous progress was triggered by exploiting novel techniques such as jet substructure and shower deconstruction~\cite{Butterworth:2008iy,Soper:2011cr,Soper:2014rya}. Today these techniques are applied by the ATLAS and CMS collaborations in experimental analyses of Higgs pair production~\cite{Sirunyan:2017isc,Aaboud:2018knk}. In our search for $hh\chi\chi$ production with four $b$-jets and missing energy, we will combine jet substructure techniques with a state-of-the-art multivariate analysis to optimize the sensitivity to our signal. For Higgs pair production in the SM, the channel with four $b$-jets does not have the best performance, due to an immense multi-jet background. In contrast, due to the presence of large missing energy in our signal, the largest background arises from electroweak gauge bosons plus jets, which lies a few orders of magnitude below the QCD multi-jet processes that appear in SM di-Higgs searches. We therefore focus on the four-bottom final state, leaving other decay channels for future exploration.

Our article is organized as follows. In Section~\ref{sec:models}, we introduce simplified models for $hh\chi\chi$ production. In Section~\ref{sec:signature}, we discuss the main features of the di-Higgs plus $\met$ signature in our simplified models and the challenges we face in reconstructing the four-bottom final state, as well as triggering and backgrounds. We attempt a cut-based analysis and investigate its discovery prospects at the High Luminosity LHC (HL-LHC). In Section~\ref{sec:mva}, we explain the details of our multi-variate analysis and demonstrate a large gain in sensitivity compared to the cut-based analysis. We stress that in the context of our simplified model, new physics could be discovered first in the di-Higgs plus $\met$ channel, while being consistent with all existing (and future HL-)LHC searches. This highlights the importance of carrying out the proposed analysis. In Section~\ref{sec:dm}, we explore the validity of the dark matter interpretation of our model. We defer our conclusions to Section~\ref{sec:conclusions}.

%% file: models.tex
\noindent In this section, we provide details on the two simplified models that give rise to the di-Higgs plus $\met$ signature, but with different final-state topologies. In the symmetric topology, the two Higgs bosons stem from a chain decay within the hidden sector, while in the resonant topology they form a di-Higgs resonance. Since the two setups have some structural similarities, we first discuss their common features. We then move on to describe the specific ingredients of the models that lead to the different topologies. 

Both models rely on an extended scalar sector with three new real scalar particles $A$, $B$, and $\chi$ that are singlets under the SM gauge group, with a mass hierarchy $m_B \gg m_A \gg m_\chi$. The models also feature a discrete $\mathbb{Z}_2$ symmetry under which particles belonging to the hidden sector are odd, while new particles in the visible sector as well as the SM particles are even. For the sake of simplicity, we assume that none of the new scalars develops a vacuum expectation value (VEV). 

The heaviest of the three new scalars, $B$, is produced via gluon fusion at the LHC and predominantly decays to $AA$ pairs. The relevant interaction terms are
\beq
\mathcal{L} \supset \frac{C_{Bgg}}{\Lambda} B G_{\mu\nu}^a G^{\mu\nu\,a} + \frac{m_{BAA}}{2} BAA.
\eeq
Here we introduce an effective dimension-five interaction of the scalar $B$ with gluons such that it is resonantly produced via $gg\to B$, in analogy to the dominant Higgs production channel in the SM. We discuss a renormalizable UV completion for this interaction in Appendix~\ref{app:UV}.

The triple scalar coupling $m_{BAA}$ induces the decay $B\to AA$ with a branching ratio near 100\%, unless $m_{BAA}$ is very small compared with the Higgs VEV $v$. For values of $C_{Bgg}$ originating from perturbative physics around the TeV scale, the decay into dijets via $B\to gg$ then occurs only at the percent level. We also suppress the decays $B \to \chi\chi$, $B \to A \chi$, and $B\to hh$ by assuming the relevant couplings to be small. Note that a $Bhh$ coupling would induce $B$-$h$ mixing, which is severely bounded by measurements of the Higgs coupling strength~\cite{Biekotter:2018rhp}. As $B$ is produced in gluon fusion, it necessarily belongs to the visible sector, i.\,e., it is even under the $\mathbb{Z}_2$ symmetry. The scalar $\chi$ belongs to the hidden sector and is thus taken to be $\mathbb{Z}_2$-odd. As it is the lightest hidden particle, it is stable and appears as missing energy in the LHC detectors.

\begin{figure}
\centering{
\begin{minipage}{.3\textwidth}
\includegraphics[width=\textwidth]{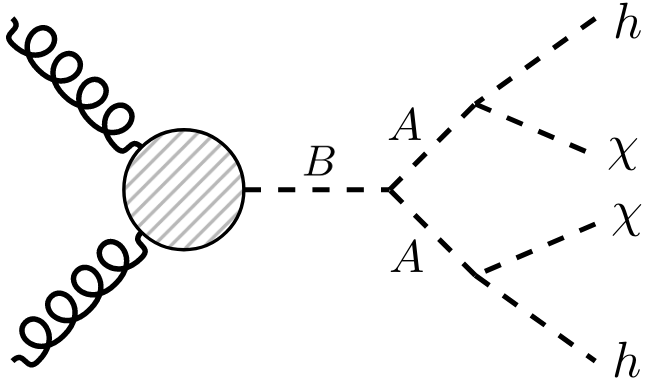}
\begin{center}
symmetric topology
\end{center}
\end{minipage}
\hspace*{2cm}
\begin{minipage}{.3\textwidth}
\includegraphics[width=\textwidth]{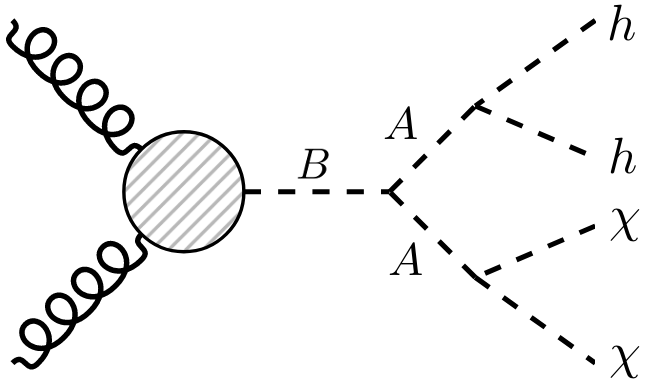}
\begin{center}
resonant topology
\end{center}
\end{minipage}}
\caption{\label{fig:topAB}Topologies for a scalar $s$-channel resonance $B$ decaying into di-Higgs plus $\met$ via a pair of scalars $AA$. If $A$ decays via $A\to h\chi$, then the symmetric topology shown on the left emerges. The decays $A\to hh$ and $A\to \chi\chi$, on the other hand, lead to the resonant topology on the right.}
\end{figure}

 Depending on the $\mathbb{Z}_2$ parity of $A$, two different event topologies for di-Higgs plus $\met$ can be distinguished,
\begin{itemize}
\item
{\bf Symmetric topology.}
If $A$ is part of the hidden sector, i.e., $\mathbb{Z}_2$-odd, it decays via $A\to h\chi$. Di-Higgs plus $\met$ arises from a symmetric event topology with chain decay in the hidden sector.
\item
{\bf Resonant topology.}
If $A$ instead belongs to the visible sector, i.e., if it is $\mathbb{Z}_2$-even, it can decay via $A \to hh$ and $A\to\chi\chi$. The di-Higgs plus $\met$ signature then arises from an asymmetric event topology and features a di-Higgs resonance.
\end{itemize}
Figure~\ref{fig:topAB} shows the event topologies for these two cases, which we now discuss in more detail.

\subsection{Symmetric topology}
\noindent In this model, both $A$ and $\chi$ are odd under the discrete $\mathbb{Z}_2$ symmetry, i.\,e.\ they belong to the hidden sector. The interaction~\footnote{Here $H$ denotes the SM Higgs doublet and gauge-invariant field contractions are assumed.}
\beq\label{eq:int-s}
\mathcal{L}_S \supset \lambda_{A\chi HH} A \chi H^\dagger H
\eeq
induces the decay $A\to h\chi$ with a branching ratio of $100\%$. Due to their different $\mathbb{Z}_2$ parity, neither $A$ nor $\chi$ can mix with the Higgs boson and thus remain pure singlets under the SM gauge group. In particular, they do not couple to electroweak gauge bosons. The coupling $\lambda_{A\chi HH}$ induces mixing of $A$ and $\chi$ after electroweak symmetry breaking. However, also the mass term $m_{A\chi}^2 A \chi$ contributes to this mixing. We conclude that the $A$-$\chi$ mixing and the decay $A\to h\chi$ are governed by two independent parameters, so that the mixing can be set to zero without affecting the $A\to h\chi$ decay in our signature. Since $B$ predominantly decays into $AA$, and $A$ decays exclusively into $h\chi$, the process
\begin{equation}
pp\to B \to AA \to(h\chi)(h\chi)
\end{equation}
is the main discovery channel for the symmetric topology at the LHC.

 This model encompasses the well studied case of electroweakino production in the MSSM, with $A$ and $\chi$ corresponding to fermions (for instance, Higgsino production as in Ref.~\cite{Matchev:1999ft}). A renormalizable coupling connecting $A$, $\chi$ and the SM Higgs implies that at least one of the fermions is non-trivially charged under the SU(2) electroweak group. Hence new states with electric charges appear, which often provide the leading collider signatures for these scenarios: multi-leptons plus $\met$ (see e.g. Ref.~\cite{Bramante:2015una}), mono-jet plus soft leptons (see e.g. Refs.~\cite{Giudice:2010wb,Schwaller:2013baa,Low:2014cba,Barducci:2015ffa}), disappearing tracks (for recent work see e.g. Refs.~\cite{Mahbubani:2017gjh,Fukuda:2017jmk,Filimonova:2018qdc,Saito:2019rtg}).

\subsection{Resonant topology}
\noindent In this model, both $A$ and $B$ are even under the $\mathbb{Z}_2$ parity, while $\chi$ is odd. The coupling $\lambda_{A\chi HH}$ is therefore forbidden and the decay $A\to h\chi$ is absent. Instead we introduce the couplings
\beq
\mathcal{L}_R \supset m_{A HH} A H^\dagger H + \frac{m_{A\chi\chi}}{2}  A\chi\chi .
\eeq 
Both of these couplings were forbidden with the symmetry assignment leading to the symmetric topology. The coupling term $m_{A HH}$ induces a mixing of $A$ with $h$ after electroweak symmetry breaking. Unlike $A$-$\chi$ mixing in the symmetric model, $A$-$h$ mixing is unavoidable here: it is induced by the same parameter as the decay $A \to hh$ that is part of the di-Higgs plus $\met$ signature. As a consequence, $A$ inherits the couplings of $h$, inducing $A\to WW$ and $A\to ZZ$ as relevant decay modes. In the limit $m_A\gg m_h,m_W,m_Z$, the decay rates fulfill the simple relation
\beq
\Gamma(A\to WW) = 2 \Gamma(A\to ZZ) = 2 \Gamma(A\to hh).
\eeq
The decay $A\to\chi\chi$ is instead induced by the coupling $m_{A\chi\chi}$, so that its decay rate can be treated as an independent parameter. To maximize the significance of the signature
\begin{equation}
pp\to B \to AA \to(hh)(\chi\chi),
\end{equation}
we assume a branching ratio of $\mathcal{B}(A\to\chi\chi) = 0.5$. In addition to the di-Higgs plus $\met$ signature, this model also gives rise to signatures with di-boson resonances and signatures with four electroweak bosons $V,V'=W,Z,h$ forming two di-boson resonances,
\begin{align}
pp \to B \to AA \to (VV)(\chi\chi), \qquad pp \to B \to AA \to (VV)(V'V').
\end{align}
These signatures and di-Higgs plus $\met$ typically occur at similar rates. They complement each other in the search for scalar hidden sectors of this kind.
\\

In summary, the two simplified models can be conveniently described by the interaction Lagrangian
\begin{equation}\label{eq:L}
\mathcal{L} = \frac{C_{Bgg}}{\Lambda} B G_{\mu\nu}^a G^{\mu\nu\,a} + \frac{m_{BAA}}{2} BAA + m_{AHH} A H^\dagger H + \frac{m_{A\chi\chi}}{2} A\chi\chi + \lambda_{A\chi HH} A \chi H^\dagger H.
\end{equation}
In the symmetric topology, the couplings $m_{AHH}$ and $m_{A\chi\chi}$ are equal to zero, due to the $\mathbb{Z}_2$ parity under which both $A$ and $\chi$ are odd. In the resonant topology, on the other hand, $\lambda_{A\chi HH} = 0$, as only $\chi$ belongs to the hidden sector. Based on the field content and symmetries of both models, additional terms could be added to the respective Lagrangians. Along this work, we will only consider those that are relevant for the collider phenomenology of the di-Higgs plus $\met$ signature
~\footnote{
In general, the UV completion for $C_{Bgg}$ can induce an additional dimension-5
operator $C_{B \gamma \gamma}$, which gives rise to a di-photon resonance in the final state. 
Whether or not $C_{B \gamma \gamma}$ is correlated with $C_{Bgg}$ depends on the specific UV completion. 
Colored and electrically neutral particles contribute to $C_{Bgg}$, but not to $C_{B \gamma \gamma}$. 
Additional vector-like leptons affect $C_{B \gamma \gamma}$, but not $C_{B gg}$. 
In the UV completion described in Appendix~\ref{app:UV}, $C_{B \gamma \gamma}$ gives rise to a branching ratio 
$\mathcal{B} (B \to \gamma\gamma)$  of a few permille for all our benchmark points. 
We thus neglect $C_{B \gamma \gamma}$ in our analysis, as well as operators such as $C_{A g g}$ and $C_{A \gamma \gamma}$ (which would appear in the resonant model, but not in the symmetric one).
}. 
Using Eq.~\eqref{eq:L}, we have implemented both Lagrangians into {\tt FeynRules}~\cite{Alloul:2013bka} and used the {\tt Universal FeynRules Output} (UFO)~\cite{Degrande:2011ua} for event generation. 

Throughout our analysis, we fix the parameters in the hidden sector as follows,
\beq
C_{Bgg} = 2.1 \cdot 10^{-3},\qquad \Lambda = 1\,\text{TeV},\qquad
m_{BAA} = v = 246\gev,\qquad
m_{A\chi\chi} \approx 2 m_{AHH}.
\eeq
The value of $C_{Bgg}$ is motivated by a UV completion with a vector-like quark with mass $m_Q = \Lambda = 1\,\text{TeV}$ and scalar coupling $y_Q = 1$, as discussed in Appendix~\ref{app:UV}. In the resonant model, the relation between $m_{AHH}$ and $m_{A\chi\chi}$ ensures a $50\%$ decay of $A\to \chi\chi$ in the limit $m_A \gg m_h,\, m_\chi$. The magnitude of these couplings does not affect the signal rate. In the symmetric model, the branching ratio of $A\to h\chi$ is 100\%, regardless of the size of the coupling $\lambda_{A\chi HH}$.

%% file: signature.tex
\noindent In order to develop a search strategy for di-Higgs plus $\met$ at the LHC, we first analyze the kinematic features of this signature and their parameter dependence in each simplified model.
We assume that $B$ is resonantly produced. 
The signal rate is then well approximated by
\begin{align}\label{eq:signal-rates}
\text{symmetric:}\quad & \sigma_S(pp\to b\bar{b}\,b\bar{b}\,\chi\chi) = \sigma(pp\to B)\times \mathcal{B}(B\to AA)\times \mathcal{B}^2(A\to h\chi)\times \mathcal{B}^2(h\to b\bar{b}),\\\nonumber
\text{resonant:}\quad & \sigma_R(pp\to b\bar{b}\,b\bar{b}\,\chi\chi) = \sigma(pp\to B)\times \mathcal{B}(B\to AA) \times 2\ \mathcal{B}(A\to hh)\, \mathcal{B}(A\to\chi\chi)\times \mathcal{B}^2(h\to b\bar{b}).
\end{align}
The couplings of $B$ do not affect the decay kinematics. Unless the $AA$ pair is produced near threshold, the heavy scalar $B$ decays almost fully via $B\to AA$, with a branching ratio of $\mathcal{B}(B\to AA) \approx 1$. Away from the threshold, the production rate for $pp\to B \to AA$ thus
depends only on the mass $m_B$ and the coupling $C_{Bgg}$. For fixed $m_B$, $m_A$, and $C_{Bgg}$, the number of produced $AA$ pairs is the same in both simplified models. Concerning the decays of $A$, in the symmetric topology $\mathcal{B}(A\to h\chi) = 1$, while in the resonant topology the maximal decay rate into $hh\,\chi\chi$ is obtained for $\mathcal{B}(A\to \chi\chi) = 0.5$. Taking into account the possible decays of $A$ into pairs of gauge bosons (see Section~\ref{sec:models}), this corresponds to $\mathcal{B}(A\to hh) \approx 0.125$. The signal rate in the symmetric model is thus about 8 times higher than in the resonant model,
\begin{align}
\sigma_S(pp\to b\bar{b}\,b\bar{b}\,\chi\chi) = \sigma(pp\to B)\times \mathcal{B}^2(h\to b\bar{b}) \approx 8\,\sigma_R(pp\to b\bar{b}\,b\bar{b}\,\chi\chi).
\end{align}
The production cross section $\sigma(pp\to B)$ can be obtained by rescaling the SM Higgs production cross section, as described in Appendix~\ref{app:UV}.

For our numerical analysis, we generate signal events at the parton level with \texttt{MG5\_aMC@NLO 2.6.1}~\cite{Alwall:2014hca}, using the \texttt{NNPDF30\_lo\_as\_0118\_nf\_4}~\cite{Ball:2014uwa} parton distribution functions implemented in the \texttt{LHAPDF 6.1.6}~\cite{Buckley:2014ana} interface. We employ \texttt{Pythia 8.2}~\cite{Sjostrand:2014zea} for parton showering and hadronization and \texttt{Delphes 3.3.3}~\cite{deFavereau:2013fsa} for a basic detector simulation, using the default implementation of the ATLAS detector.  Crucial inputs for our analysis 
are the $b$-tagging efficiency, $\epsilon_b$, and the light and charm jet mistag rates, $\epsilon_l$ and $\epsilon_c$, which are given by the following $p_T$-dependent functions
\begin{equation}
\epsilon_b = 0.8 \frac{30 \tanh(3 \cdot 10^{-3}  p_T)}{1+ 8.6 \cdot 10^{-2}  p_T} \,, \qquad \epsilon_l = 0.002  (1+3.65 \cdot 10^{-3} p_T) \,,\qquad \epsilon_c = 0.2 \frac{\tanh(0.02 p_T)}{1+ 3.4 \cdot 10^{-3}  p_T} \, .
\end{equation}
Hence, for a jet transverse momentum of $p_T(j) = 50\,(250)\gev$, we find $\epsilon_b, \epsilon_l, \epsilon_c = 67,0.24,0.26\,(73,0.3,1.0) \%$.
\subsection{Kinematics and benchmarks}\label{sec:kinematics}
\noindent The kinematics of our signature is driven by the available phase space in the $B$ and $A$ decays. We parametrize this in either model in terms of the mass differences of the involved particles,
\begin{align}
\text{symmetric:}\quad & \Delta_{BA} = m_B - 2m_A,\qquad \Delta_{Ah\chi} = m_A - (m_h + m_\chi),\\\nonumber
\text{resonant:}\quad & \Delta_{BA} = m_B - 2m_A,\qquad \Delta_{Ah\chi} = \text{min}(m_A - 2 m_h, m_A - 2 m_\chi).
\end{align}
We fix the scalar mass $m_B = \{1000,750,500\}\gev$ and scan $\Delta_{BA}$ in steps of $75\gev$ over the kinematically accessible region. In the symmetric model, $\Delta_{Ah\chi}$ is scanned in steps of $100\gev$. We require $\Delta_{Ah\chi} > 10\gev$ to prevent too strong a phase-space suppression in $A$ decays. In the resonant model, we fix $m_\chi = 25\gev$ to satisfy the different kinematic boundaries.\footnote{As long as long as the $A\to\chi\chi$ final state is kinematically accessible, the mass $m_\chi$ has no impact on the di-Higgs plus $\met$ phenomenology in the resonant model.} The so-obtained benchmark scenarios for the symmetric ($S$) and resonant ($R$) model are labelled as
\begin{align}
S\textunderscore m_B\textunderscore m_A\textunderscore m_\chi,\qquad R\textunderscore m_B\textunderscore m_A\textunderscore m_\chi,
\end{align}
where the masses of the scalars are given in units of GeV. They are shown for both models in Table~\ref{tab:benchChain}, where we also give the corresponding signal rates, $\sigma_S$ and $\sigma_R$, as defined in Eq.~(\ref{eq:signal-rates}). We have verified that $\mathcal{B}(B\to AA) \gtrsim 97\,\%$ in both topologies for all benchmarks.
\begin{table}[]
\centering
\begin{minipage}[t]{.35\textwidth}
\begin{tabular}{|l|l|}
\hline
{\bf symmetric benchmark \# }                                                                & {\bf\boldmath$\sigma_S$ {[}fb{]}} \\ \hline\hline
$S\textunderscore 1000 \textunderscore 475 \textunderscore \{340,250,150,50\}\quad$ & 2.94                \\ \hline
$S\textunderscore 1000 \textunderscore 400 \textunderscore \{265,175,75\}$     & 2.99                \\ \hline
$S\textunderscore 1000 \textunderscore 325 \textunderscore \{190,100,0.1\}$    & 3.00                \\ \hline
$S\textunderscore 1000 \textunderscore 250 \textunderscore \{115,25\}$         & 3.01                \\ \hline
$S\textunderscore 750 \textunderscore 350 \textunderscore \{215,125,25\}$      & 15.25               \\ \hline
$S\textunderscore 750 \textunderscore 275 \textunderscore \{140,50\}$          & 15.32               \\ \hline
$S\textunderscore 750 \textunderscore 200 \textunderscore \{65\}$              & 15.33               \\ \hline
$S\textunderscore 500 \textunderscore 225 \textunderscore \{90,50,0.1\}$       & 106.10              \\ \hline
$S\textunderscore 500 \textunderscore 200 \textunderscore \{65,25\}$           & 106.12              \\ \hline
$S\textunderscore 500 \textunderscore 175 \textunderscore \{40,0.1\}$          & 106.14              \\ \hline
$S\textunderscore 500 \textunderscore 150 \textunderscore \{15\}$              & 106.15              \\ \hline
\end{tabular}
\end{minipage}
\hspace{2cm}
\begin{minipage}[t]{.35\textwidth}
\begin{tabular}{|l|l|}
\hline
{\bf resonant benchmark \#} \hspace*{.5cm}                                                & 
{\bf\boldmath$\sigma_R$ {[}fb{]}} \\ \hline\hline
$R\textunderscore 1000 \textunderscore 475 \textunderscore 25$ & 0.37                 \\ \hline
$R\textunderscore 1000 \textunderscore 450 \textunderscore 25$ & 0.38              \\ \hline
$R\textunderscore 1000 \textunderscore 400 \textunderscore 25$ & 
0.38              \\ \hline
$R\textunderscore 1000 \textunderscore 350 \textunderscore 25$ & 
0.37             \\ \hline
$R\textunderscore 1000 \textunderscore 325 \textunderscore 25$ & 
0.36             \\ \hline
$R\textunderscore 1000 \textunderscore 275 \textunderscore 25$ & 0.29              \\ \hline
$R\textunderscore 750 \textunderscore 350 \textunderscore 25$  &
1.88           \\ \hline
$R\textunderscore 750 \textunderscore 325 \textunderscore 25$  & 
1.84            \\ \hline
$R\textunderscore 750 \textunderscore 275 \textunderscore 25$  & 
1.48            \\ \hline
\end{tabular}
\end{minipage}
\caption{Benchmarks $S\textunderscore m_B \textunderscore m_A \textunderscore m_\chi$ (symmetric topology, left) and $R\textunderscore m_B \textunderscore m_A \textunderscore m_\chi$ (resonant topology, right) for a di-Higgs plus $\met$ signature. The second column shows the signal rate $\sigma_{S,R} (p p \to b\bar{b}\,b\bar{b}\,\chi\chi)$ at a centre-of-mass energy of 14 TeV, assuming a UV completion by a vector-like quark with $m_Q = 1\tev$ and $y_Q=1$, see Appendix~\ref{app:UV}.}\label{tab:benchChain}
\end{table}

In addition to the di-Higgs plus $\met$ signature, our models induce processes like $ p p \to B \to jj$ and $ p p \to B \to \gamma \gamma$, as well as mono-jet, di-jet plus $\met$, and mono-Higgs signatures. The highest sensitivity to our scenarios is expected in searches for top squarks, bottom squarks and gluinos~\cite{Aad:2016eki,Aaboud:2017ayj,Aaboud:2017vwy,ATLAS:2017dnw}, which focus on large $\met$ together with $b$-jets and light jets. Using {\tt CheckMATE2}~\cite{Dercks:2016npn}, we have verified that all benchmarks evade existing LHC searches for these and similar processes at $\sqrt{s}=8$ and $13\tev$. Current searches for $4b + \met$ lack sensitivity to our models, because they focus on phase-space regions that are sparsely populated by our signal.

We have also verified that future searches at the HL-LHC with $3\,\text{ab}^{-1}$ of data will not be competitive with our final state. Since the $14\,$TeV analyses available in {\tt CheckMATE2} are only a handful, we have estimated the reach of the remaining analyses by naively rescaling the expected reach of the current $13\tev$ studies with the square root of the luminosity. We conclude that our benchmarks are not only viable today, but also they will not be probed by future LHC searches. Di-Higgs plus $\met$ can thus be considered the discovery channel for these scenarios. From the resonant topology, we predict additional signatures with di-boson resonances and missing energy, like $WW+\met$ and $ZZ+\met$, as well as signatures with four electroweak  bosons forming two di-boson resonances, cf. Section~\ref{sec:models}. Since these signatures are expected to occur with similar rates as di-Higgs plus $\met$ production, they can serve as complementary discovery channels for the resonant topology.

\begin{figure}[t]
\centering
\includegraphics[width=0.435\textwidth]{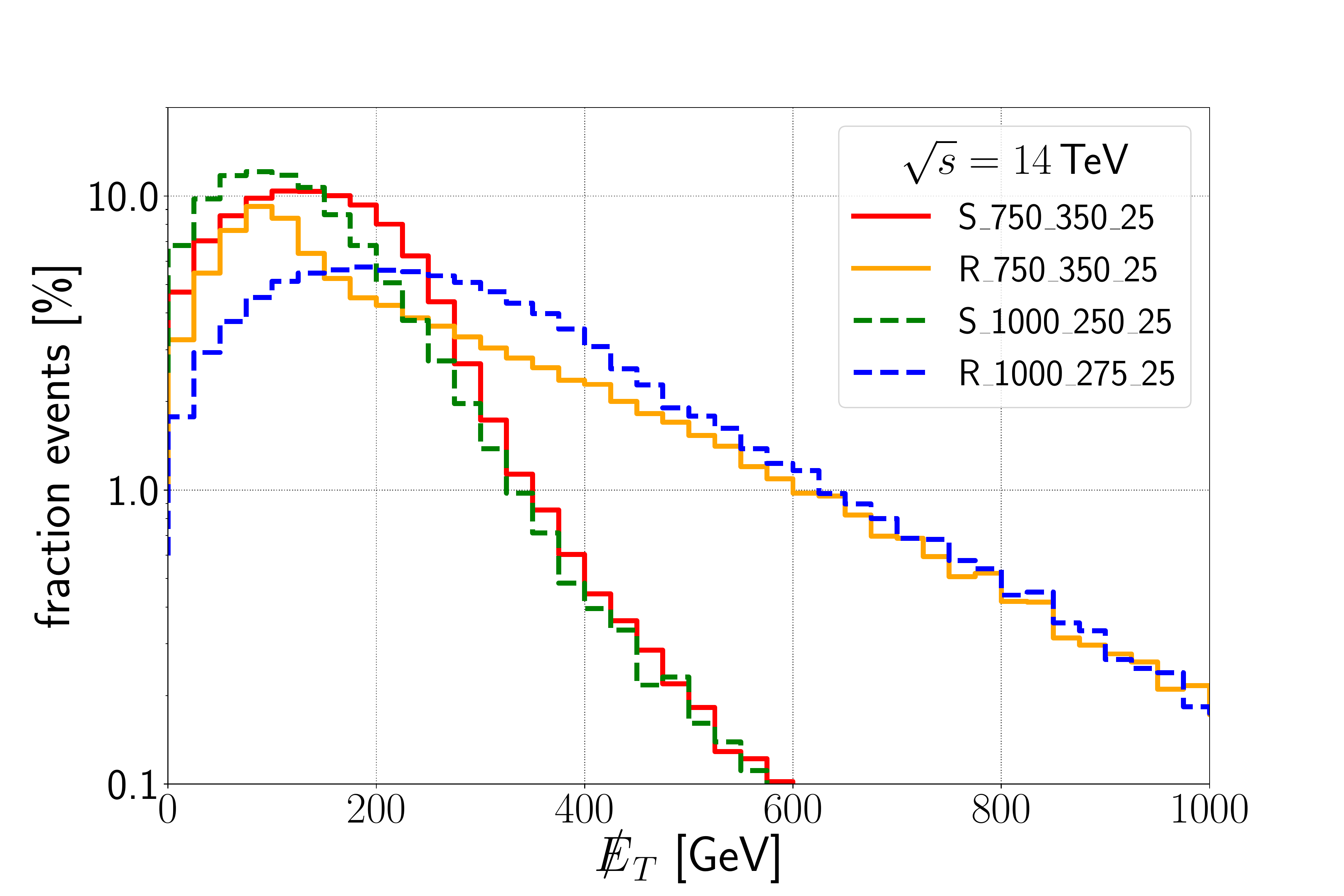} \hspace*{0.1cm}
\includegraphics[width=0.265\textwidth]{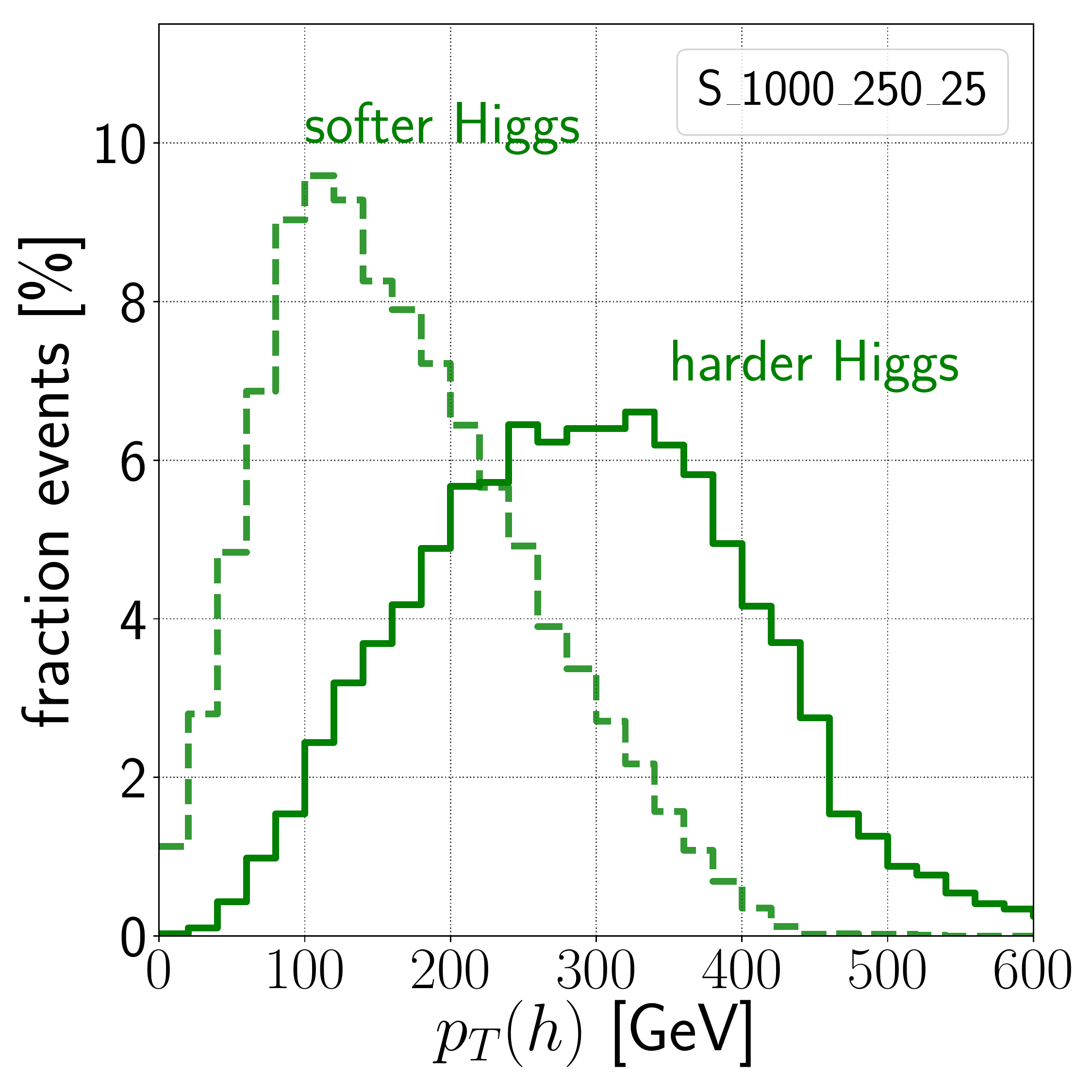} \hspace*{0.1cm}
\includegraphics[width=0.265\textwidth]{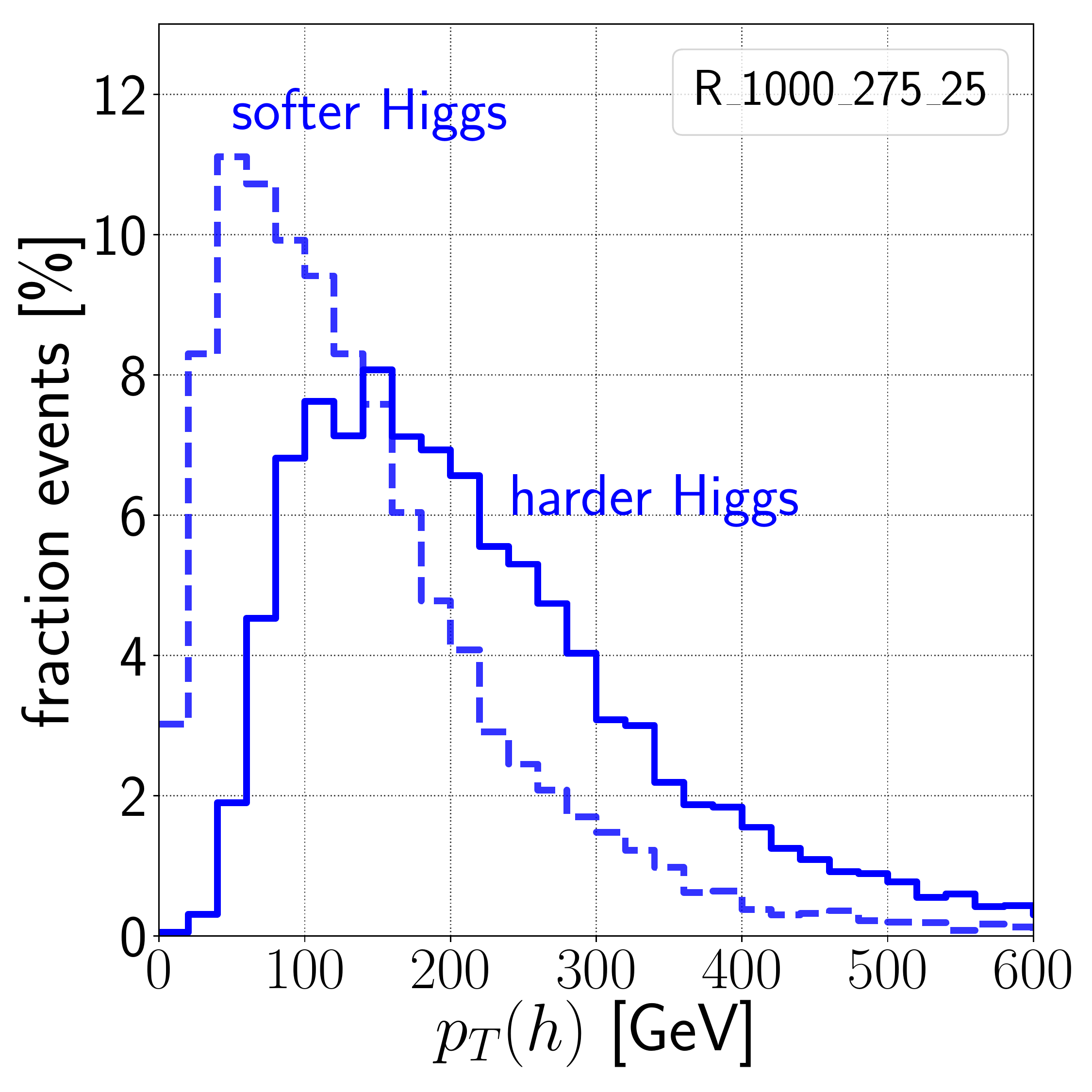}
\caption{\label{fig:met_distro_topo}Left panel: Reconstructed $\met$ distribution for four benchmarks. Solid lines correspond to a compressed spectrum, $S\textunderscore750\textunderscore350\textunderscore 25$ (red) and $R\textunderscore 750\textunderscore 350\textunderscore 25$ (orange); dashed lines represent a split spectrum, $S\textunderscore 1000\textunderscore 250\textunderscore 25$ (green) and $R\textunderscore 1000\textunderscore 275\textunderscore 25$ (blue). Center and right panels: transverse momentum distribution of the harder (solid) and the softer (dashed) Higgs bosons at the parton level, for the benchmarks $S\textunderscore 1000 \textunderscore 250 \textunderscore 25$  (center panel) and $R \textunderscore 1000 \textunderscore 275\textunderscore 25$ (right panel).}
\end{figure}

To illustrate the phenomenology of our simplified models, we use the specific benchmarks 
\begin{align}
S\textunderscore750\textunderscore350\textunderscore25,\quad R\textunderscore750\textunderscore350\textunderscore25,\qquad S\textunderscore1000\textunderscore250\textunderscore25,\quad R\textunderscore1000\textunderscore275\textunderscore25.
\end{align}
The first two benchmarks correspond to scenarios with little phase space for the $B\to AA$ decay (\emph{compressed spectrum}), the second two with large phase space (\emph{split spectrum}). In our benchmarks, a compressed (split) spectrum is parametrized by a small (large) $\Delta_{BA}$, which determines the boost of $A$. The kinematic differences between the two models originate in the respective $A$ decays, which are imprinted on the $\met$ distribution. In the left panel of Figure~\ref{fig:met_distro_topo}, we show the $\met$ distribution for the four benchmarks, illustrating the effects of a compressed and split spectrum in each topology. We also present the transverse momentum distributions of the Higgs bosons at parton level, $p_T(h)$, for a split spectrum in the symmetric topology (center panel) and the resonant topology (right panel).

In the symmetric topology, the two $A$ particles are produced back-to-back in the center-of-mass frame and split their transverse momentum into $\chi$ and $h$. The vector sum of their transverse momenta is thus subject to cancellations. The peak position of the $\met$ distribution depends on the available phase space in both the $B$ and $A$ decays, $\Delta_{BA} =  m_B - 2m_A$ and $\Delta_{Ah\chi} = m_A - (m_h + m_\chi)$. For larger values of $\met$, the distribution drops fast. In the resonant topology, the $\met$ distribution is equal to the transverse momentum distribution of $A$. The peak of the $\met$ distribution depends now only on $\Delta_{BA} =  m_B - 2m_A$, and the spectrum is harder at large $\met$ than for the symmetric topology. A trigger on missing energy will thus favor one or the other topology, depending on the position of the $\met$ cut. The transverse momentum distribution of the Higgs bosons peaks at lower momenta than the $\met$ distribution. As can be seen by inspecting the transverse momenta of the Higgs bosons, the $b$-jets from Higgs decays are less likely to pass the trigger requirements, which typically imply strong cuts on the jet transverse momentum~\cite{Aaboud:2016leb}. Triggering on missing transverse energy rather than on the $b$-jets yields a more efficient signal selection.

\subsection{Jet substructure technique}
\noindent Depending on the model and the mediator spectrum, the $b$-jets from Higgs decays can be produced with a large boost. The $b$-jets are thus collimated and cannot be resolved as individual jets. To reconstruct the boosted $h\to b \bar{b}$ decays, we crucially rely on jet substructure techniques. The current substructure module in Delphes, \texttt{SoftDrop}, is a modified version of the BDRS algorithm~\cite{Butterworth:2008iy} that includes $b$-tagging and flavor tagging for fat jets. To make the tool applicable for our purposes, we have extended these functionalities to subjets. Based on \texttt{SoftDrop}, we have developed two new modules called \texttt{JetFlavorAssociationSubjets} and \texttt{BTaggingSubjets}.~\footnote{The corresponding code can be obtained from the authors upon request.} These modules allow us to access the four-momenta and $b$-tags of each fat jet in the event, and also of each subjet associated to it in the Delphes output. We will  speak of ``$x$-$y$ $b$-tags'' to describe an event selection where one fat jet contains at least $x$ $b$-tagged subjets and another fat jet contains at least $y$ $b$-tagged subjets. The performance of our tagging technique depends on the fat-jet radius, $R$. We use $R=1.2$ for the symmetric and $R=0.6$ for the resonant topology.~\footnote{The choice of these values and their impact on the analysis will be explained at the end of this subsection.}

\begin{figure}[t]
\centering
\includegraphics[height=2.5in]{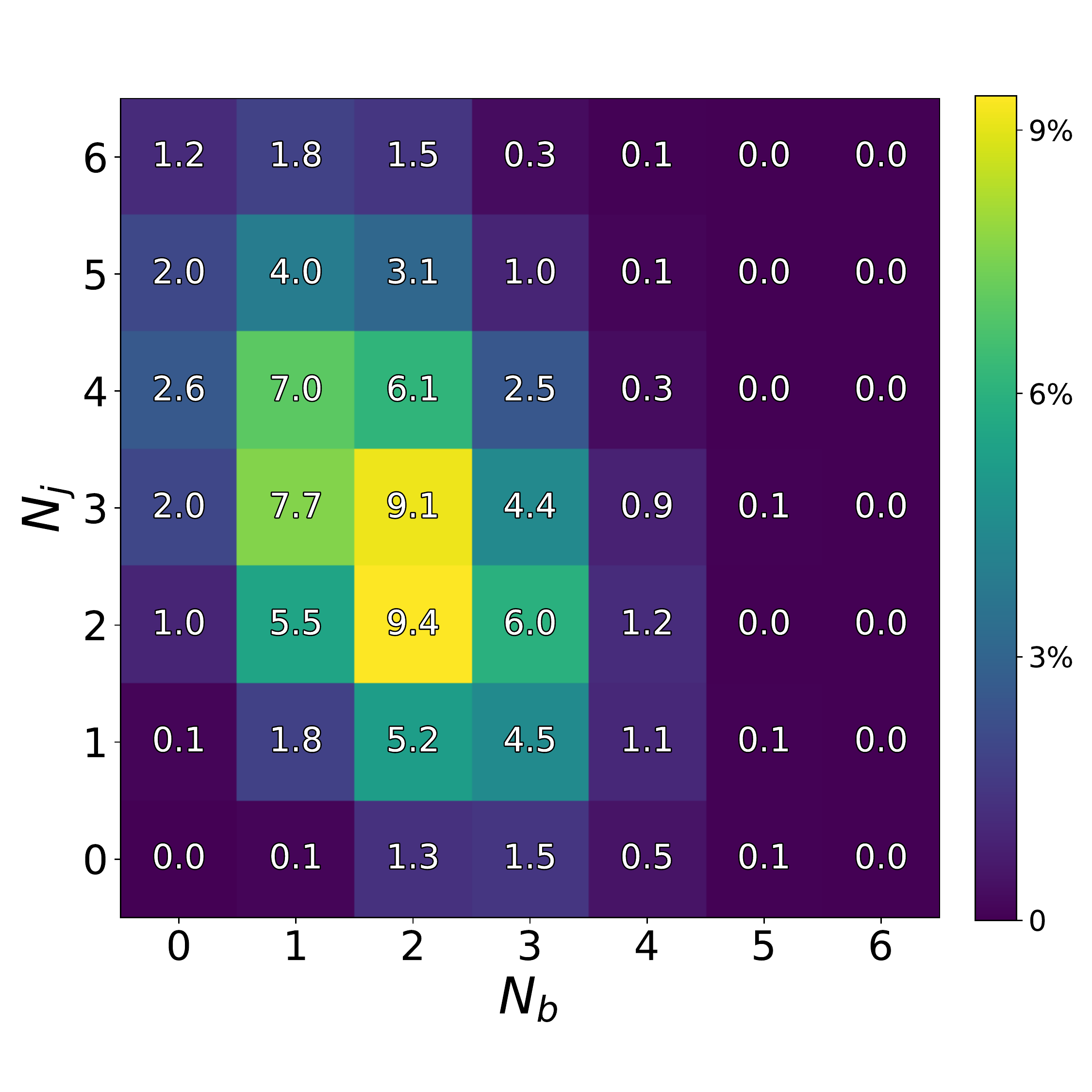}
\caption{\label{fig:where-bs-go}Signal distribution of the number of reconstructed $b$-jets, $N_b$, and the number of reconstructed light jets, $N_j$, for the benchmark $R\textunderscore1000\textunderscore275\textunderscore25$. Shown is the fraction of signal events in percent.
}
\end{figure}

\begin{figure}[t]
\centering
\includegraphics[height=2.5in,width=0.3\textwidth]{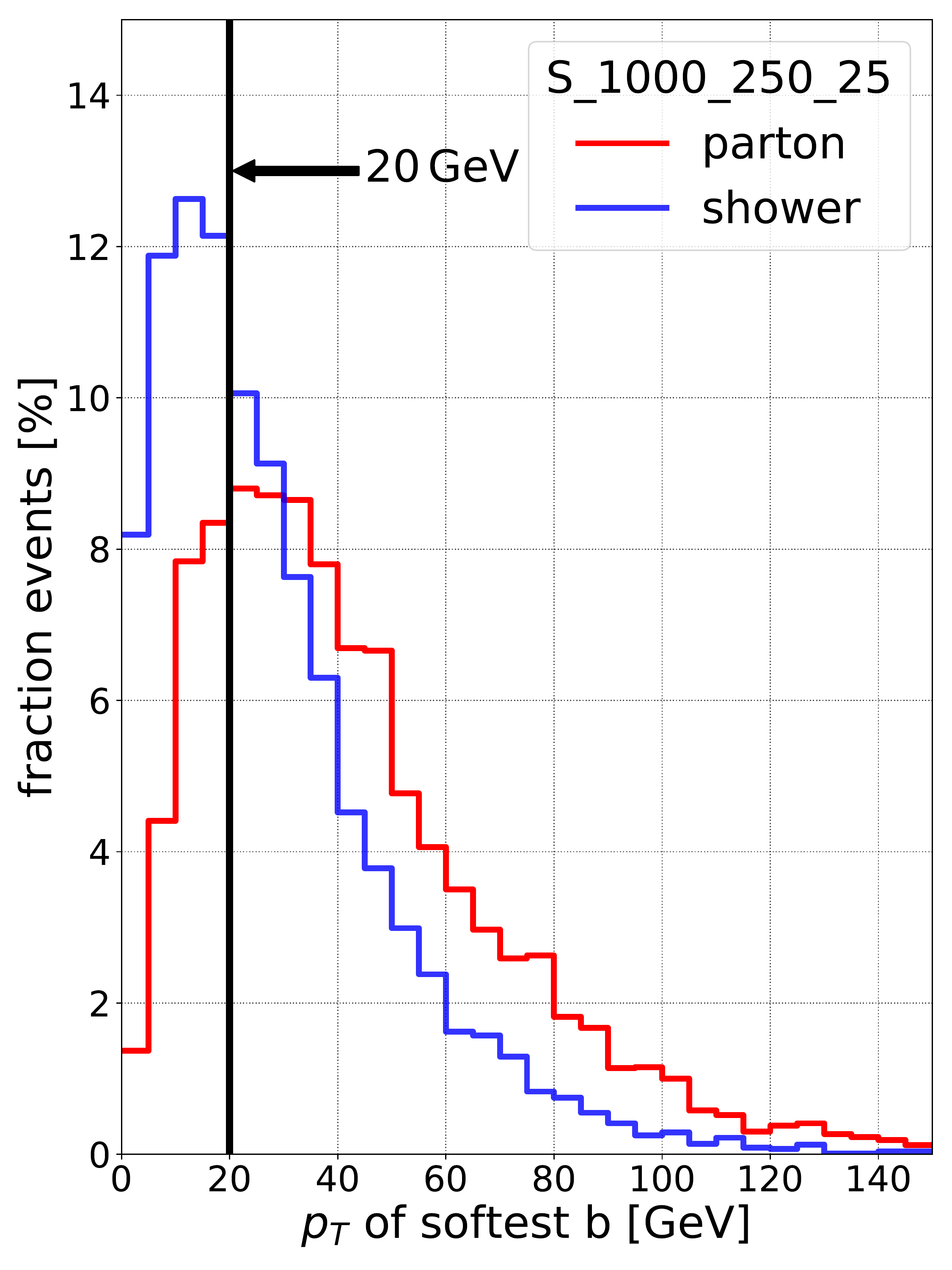} \hspace*{0.5cm}
\includegraphics[height=2.5in,width=0.3\textwidth]{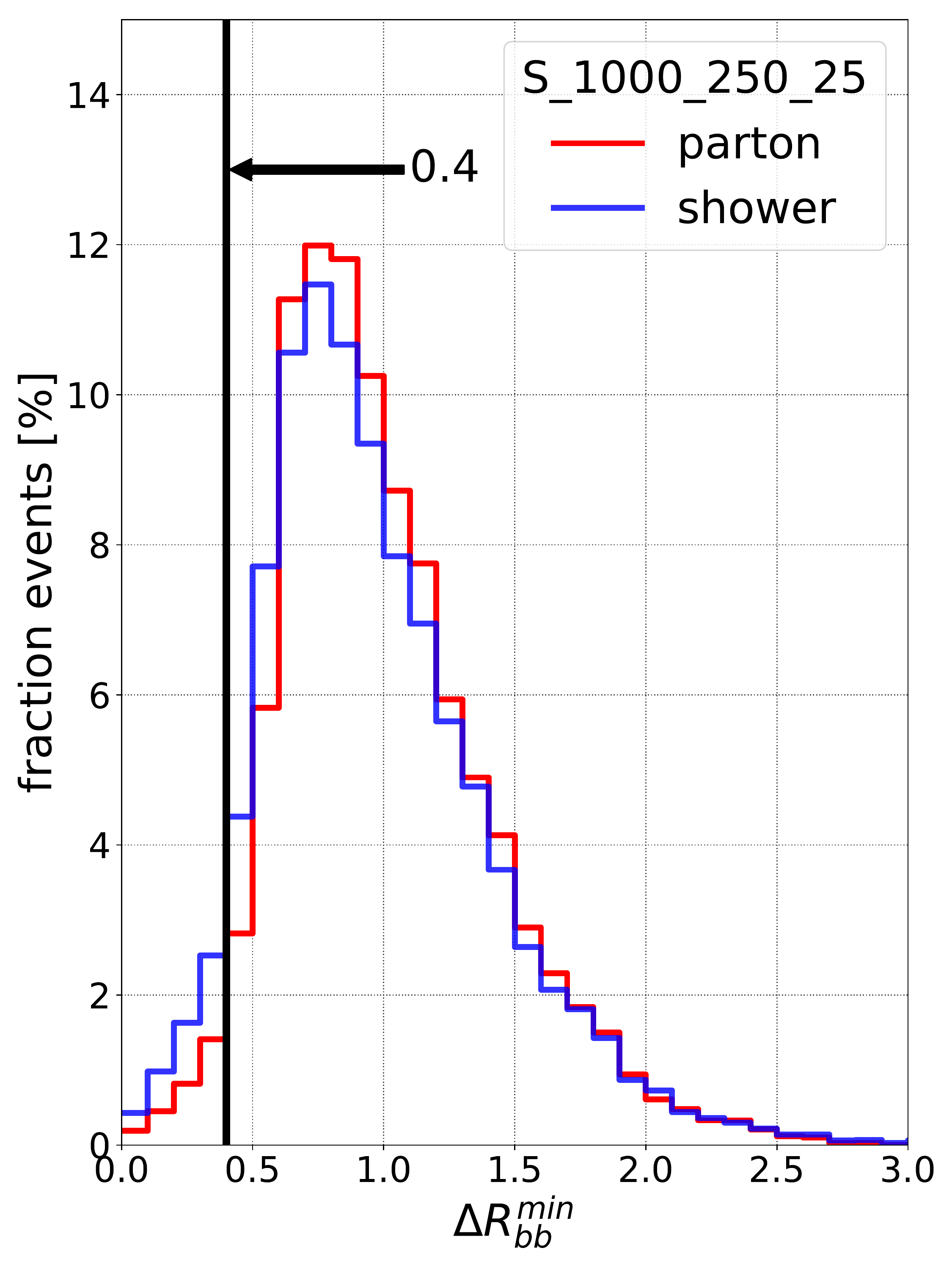} \hspace*{0.5cm}
\includegraphics[height=2.5in,width=0.3\textwidth]{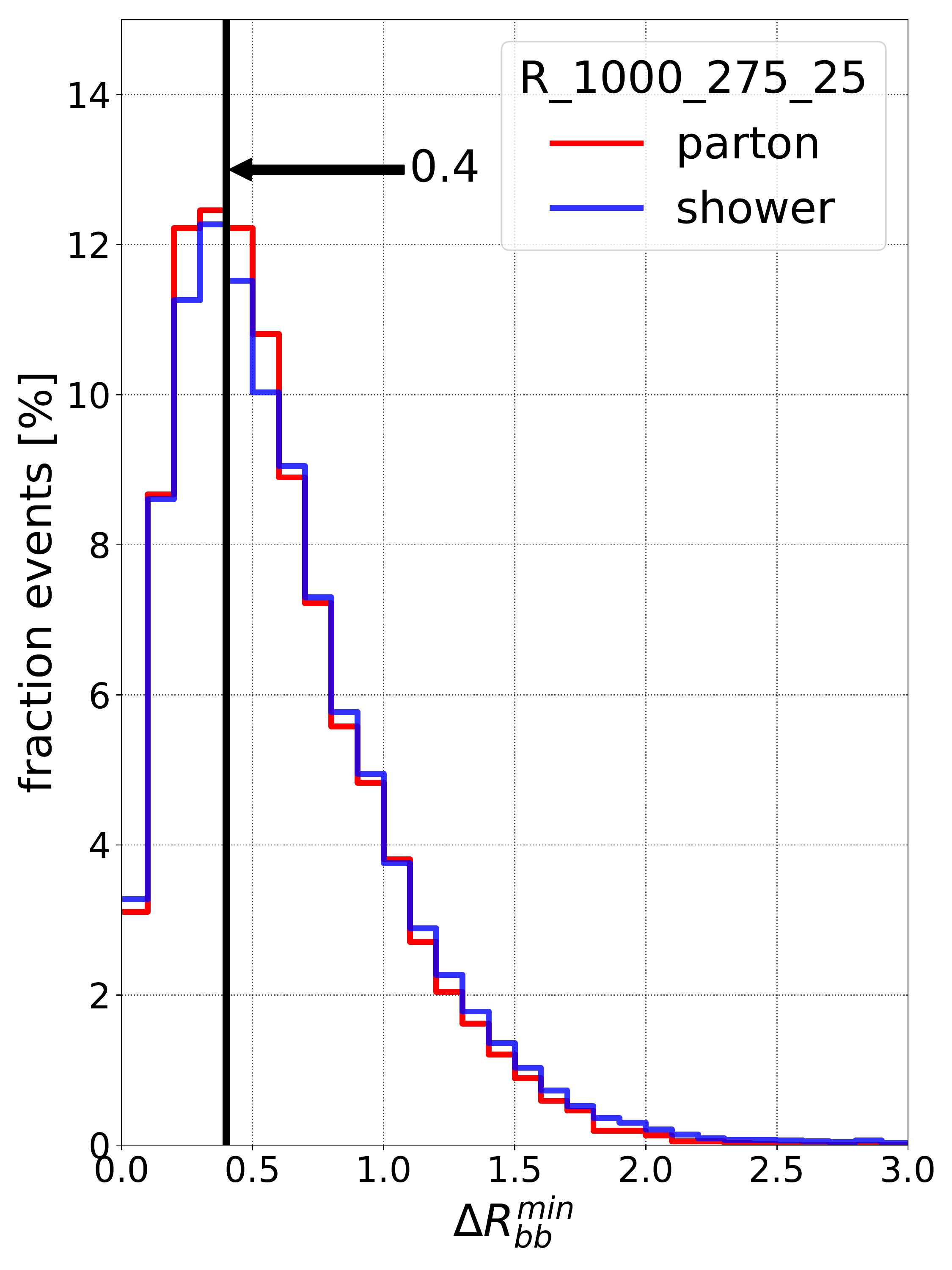} \hspace*{0.5cm}
\caption{\label{fig:soft-collinear_bs}Transverse momentum distribution of the softest $b$-quark (left panel) and minimum distance between any pair of $b$-quarks ($\Delta R_{bb}^{\rm min}$) in an event (center panel) for the benchmark $S\textunderscore1000\textunderscore250\textunderscore25$, at the parton level (red) and after showering (blue). Right panel: $\Delta R_{bb}^{\rm min}$, as in the center panel, but for $R\textunderscore1000\textunderscore275\textunderscore25$.}
\end{figure}

Due to the limited $b$-tagging efficiency and rejection efficiency of light (i.e., non-$b$-tagged) jets, as well as the jet rapidity cut of $|\eta_b| < 2.5$, not all of the four $b$-subjets in our signal will be tagged. To quantify this statement, we show in Figure~\ref{fig:where-bs-go} the number of $b$-tagged subjets, $N_b$, versus the number of light jets, $N_j$, for an exemplary benchmark of the resonant topology, $R\textunderscore1000\textunderscore275\textunderscore25$. Other resonant benchmarks show a similar pattern, and the behavior is similar for the symmetric topology.
 It is apparent that most of the time only three or fewer $b$-subjets are reconstructed. Note, however, that the amount of missing $b$-jets is larger than the naive estimation from the plain $b$-tagging efficiency. This loss is due to a significant number of reconstructed jets that are either \emph{soft}, i.e., carry $p_T(j) < 20\gev$, and/or \emph{collinear}, i.e., with angular separation of $\Delta R_{bb} < 0.4$.~\footnote{While the detector can resolve jets and subjets with smaller angular separation, regular jets are clustered in ATLAS with a $k_T$ algorithm with $R=0.4$.} In Figure~\ref{fig:soft-collinear_bs}, we show the $p_T$ distribution of the softest $b$-quark (left panel) and the minimum distance between any pair of $b$-quarks ($\Delta R_{bb}^{\rm min}$) in an event for the symmetric benchmark $S\textunderscore1000\textunderscore250\textunderscore25$ (center panel), at the parton level (red) and after showering (blue). In the right panel of Figure~\ref{fig:soft-collinear_bs}, we show $\Delta R_{bb}^{\rm min}$ for the resonant benchmark $R\textunderscore1000\textunderscore275\textunderscore25$.
 From the left panel, we clearly see that the parton shower reduces the average transverse momentum of the $b$-quarks below the detector threshold of $20\gev$. A tight event selection with $2$-$2$ $b$-tags would cut away a large amount of signal. We therefore apply looser requirements on the $b$-tags in our analysis (see Section~\ref{subsec:cutflow}).

Now we turn our attention to the $\Delta R_{bb}^{\rm min}$ distribution. As expected, the parton shower barely changes the collinearity of the $b$-quarks. Therefore, if $b$-jets are not reconstructed as such, this is due to the characteristic mass spectrum of the model, rather than parton showering. 
The $\Delta R_{bb}^{\rm min}$ distribution depends both on the boost of the Higgs bosons and on the event topology. In a split spectrum, the Higgs bosons are more boosted, so that $b$-quarks from the same Higgs decay are closer to each other. In a compressed spectrum, the Higgs bosons are softer and the $b$-quarks are emitted with a larger angular separation. Naively one might thus think that smaller values of $\Delta R_{bb}^{\rm min}$ are preferred in the symmetric model, where the Higgs bosons carry larger transverse momenta. In Figure~\ref{fig:soft-collinear_bs}, however, we observe the opposite behaviour. 
This is due to the different event topology:  In the resonant model, the two Higgs bosons stem from the decay of one $A$ boson and are thus much closer than in the symmetric model, where they originate from opposite sides of the event. Consequently in the resonant model all four $b$-jets tend to be collimated, while in the symmetric model the two pairs of $b$-jets are well separated. We hence conclude that only in the symmetric model $\Delta R_{bb}^{\rm min}$ is a direct measure of the Higgs transverse momentum. In the resonant model, on the other hand, the closest $b$-jets do not always stem from the same Higgs decay, so that $\Delta R_{bb}^{\rm min}$ is also sensitive to the boost of $A$. In the symmetric model, the parton level requirement $\Delta R_{bb} > 0.4$ only cuts away a few percent of the signal events. In contrast, in the resonant model this cut has an important impact on the signal. This loss of events, together with the lower total event rates discussed in Section~\ref{sec:models}, suggests that the resonant topology is harder to find that the symmetric one for a given particle spectrum. The fact that the maximum of the $\Delta R_{bb}^{\rm min}$ distribution in the resonant model lies at lower values than in the symmetric model motivates different choices of fat-jet radii. We use $R=1.2$ for the symmetric and $R=0.6$ for the resonant topology.

\subsection{Backgrounds and cutflow analysis}
\label{subsec:cutflow}
\noindent The main SM backgrounds to our signal of $4b + \met$ are due to $W+\text{jets}$, $Z+\text{jets}$, as well as top-antitop production with one leptonic and one hadronic decay. All backgrounds have been simulated using {\tt Sherpa 2.2.1}~\cite{Gleisberg:2008ta} at leading order (LO) in QCD, including parton shower and hadronization effects. We use the same setup as for the signal generation, as described at the end of Section~\ref{sec:kinematics}. Our analysis has been performed with {\tt ROOT}~\cite{Brun:1997pa}.

\begin{figure}[t]
\begin{minipage}{.34\textwidth}
\includegraphics[height=2in,width=1.19\textwidth]{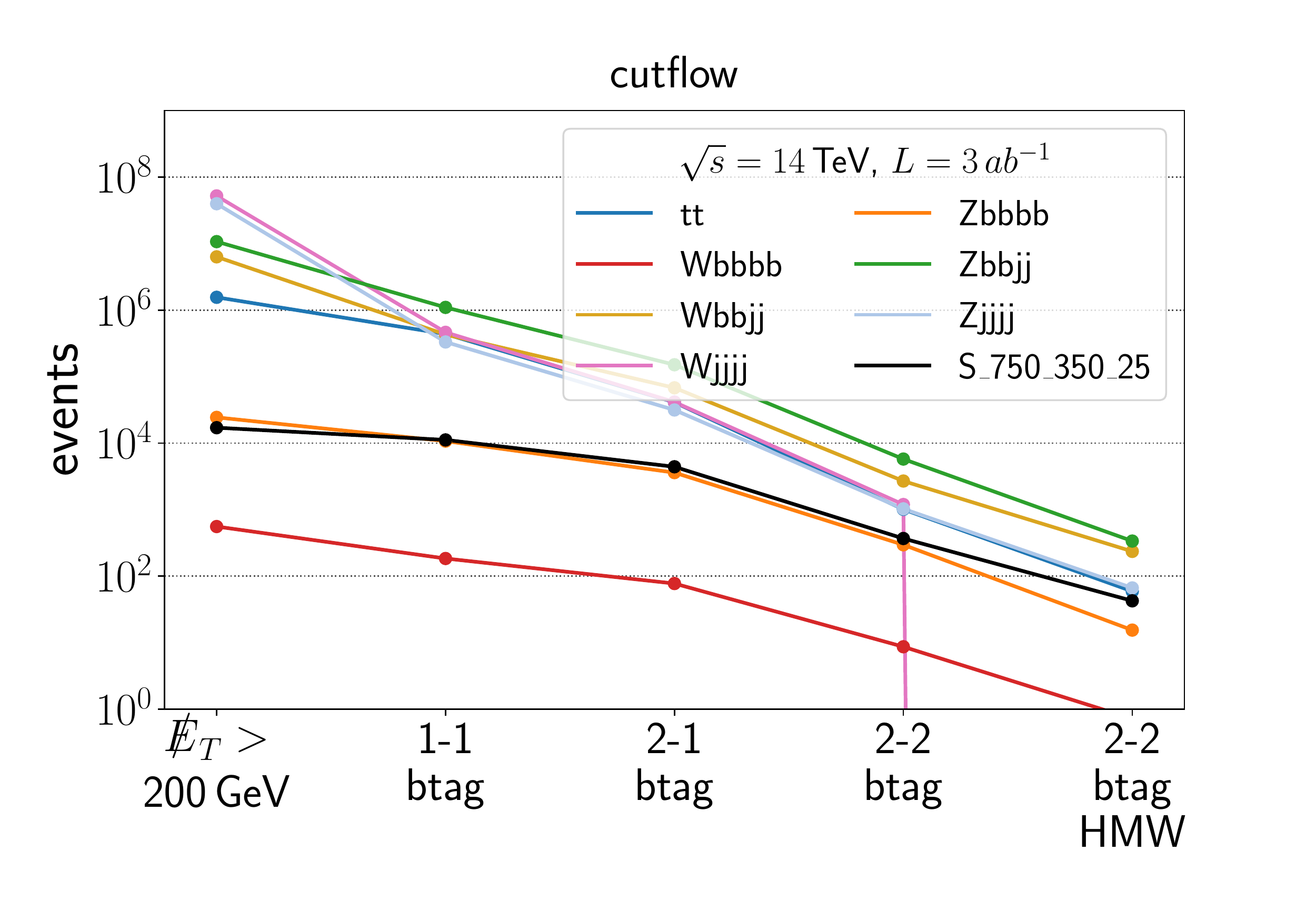}
\end{minipage} \hfill
\begin{minipage}{.34\textwidth}
\includegraphics[height=2in,width=1.17\textwidth]{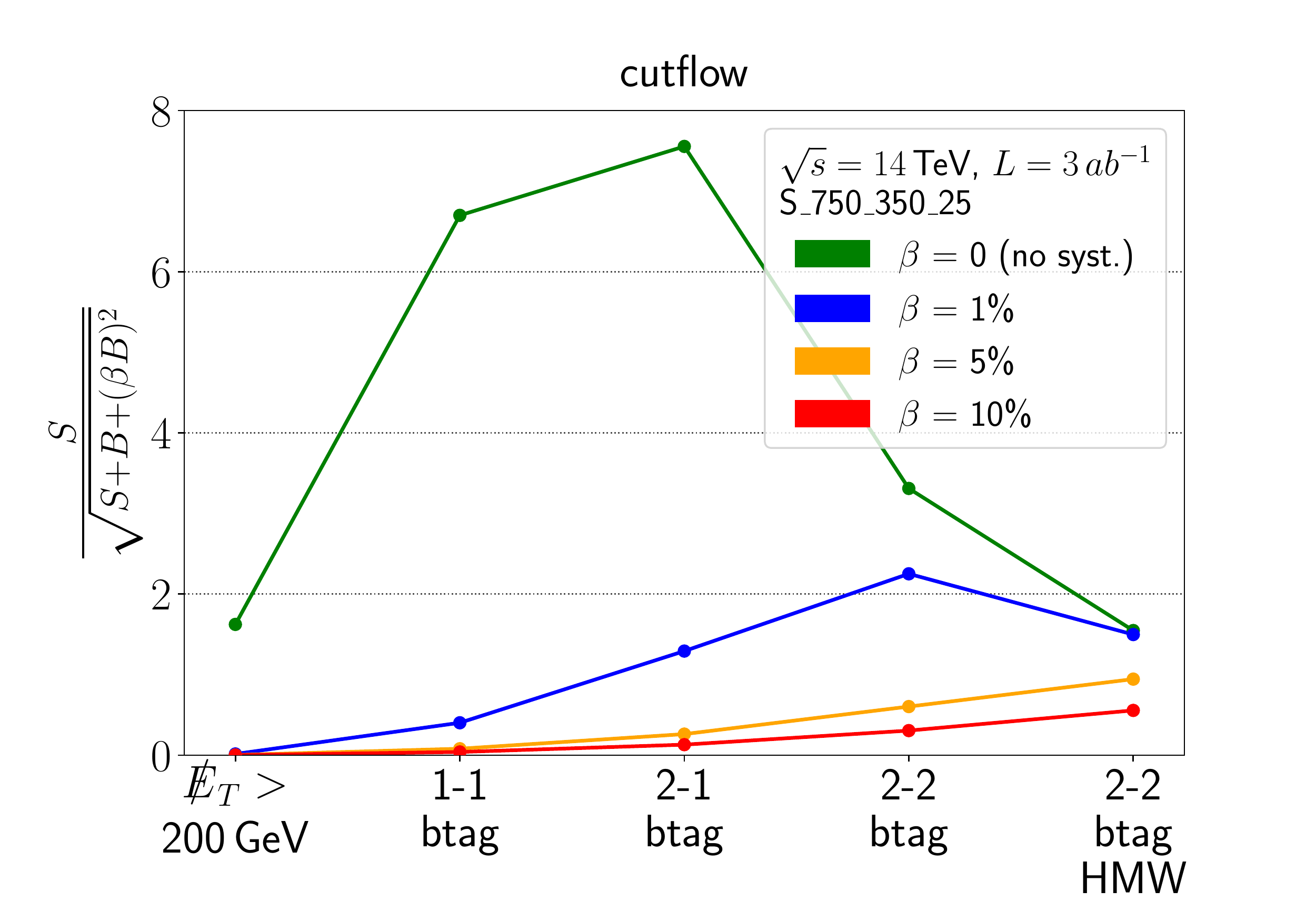}
\end{minipage} \hfill
\begin{minipage}{.26\textwidth}
\vspace*{-0.30cm}
 \begin{tabular}{|l|c|}
 \hline
  {\bf backgd. ($R=1.2$)} & {\bf\boldmath $\sigma$ [fb]} \\\hline
  \hline
  $Z(\rightarrow\nu\bar{\nu})\,b\bar{b}jj$ & 366 \\\hline
  $W(\rightarrow\ell\nu)\,jjjj$ & 154 \\\hline
  $t\,(\to bjj)\,\bar{t}\,(\to \bar{b}\ell\nu)$  & 146\\\hline
  $W(\rightarrow\ell\nu)\,b\bar{b}jj$ & 143 \\\hline
  $Z(\rightarrow\nu\bar{\nu})\,jjjj$ & 110 \\\hline
  $Z(\rightarrow\nu\bar{\nu})\,b\bar{b}b\bar{b}$ & 4 \\
\hline
 \end{tabular}
\end{minipage}
\caption{\label{fig:cutflowChain}Cutflow for number of events (left panel) and significance $\Sigma$ (center panel) for  
the cut-based analysis of the symmetric model $S\textunderscore750\textunderscore350\textunderscore25$. Right panel: Cross section of dominant backgrounds for $\met > 200\gev$, after applying a lepton veto and requiring 1-1 $b$-tagging, i.e., at least one $b$-tagged subjet from each fat jet with radius $R=1.2$.}
\end{figure}

\begin{figure}[t]
\begin{minipage}{.34\textwidth}
\includegraphics[height=2in,width=1.19\textwidth]{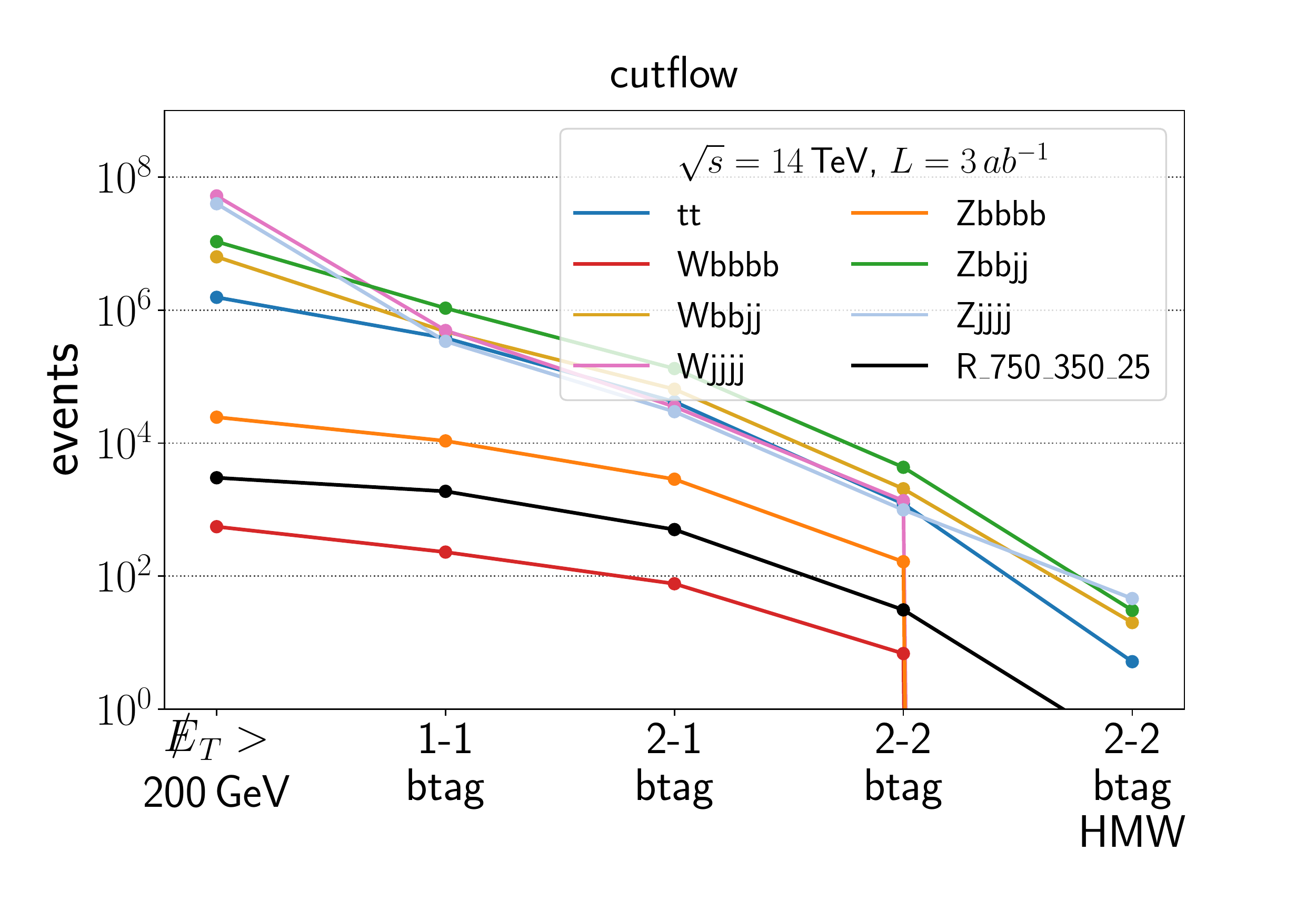}
\end{minipage} \hfill
\begin{minipage}{.34\textwidth}
\includegraphics[height=2in,width=1.17\textwidth]{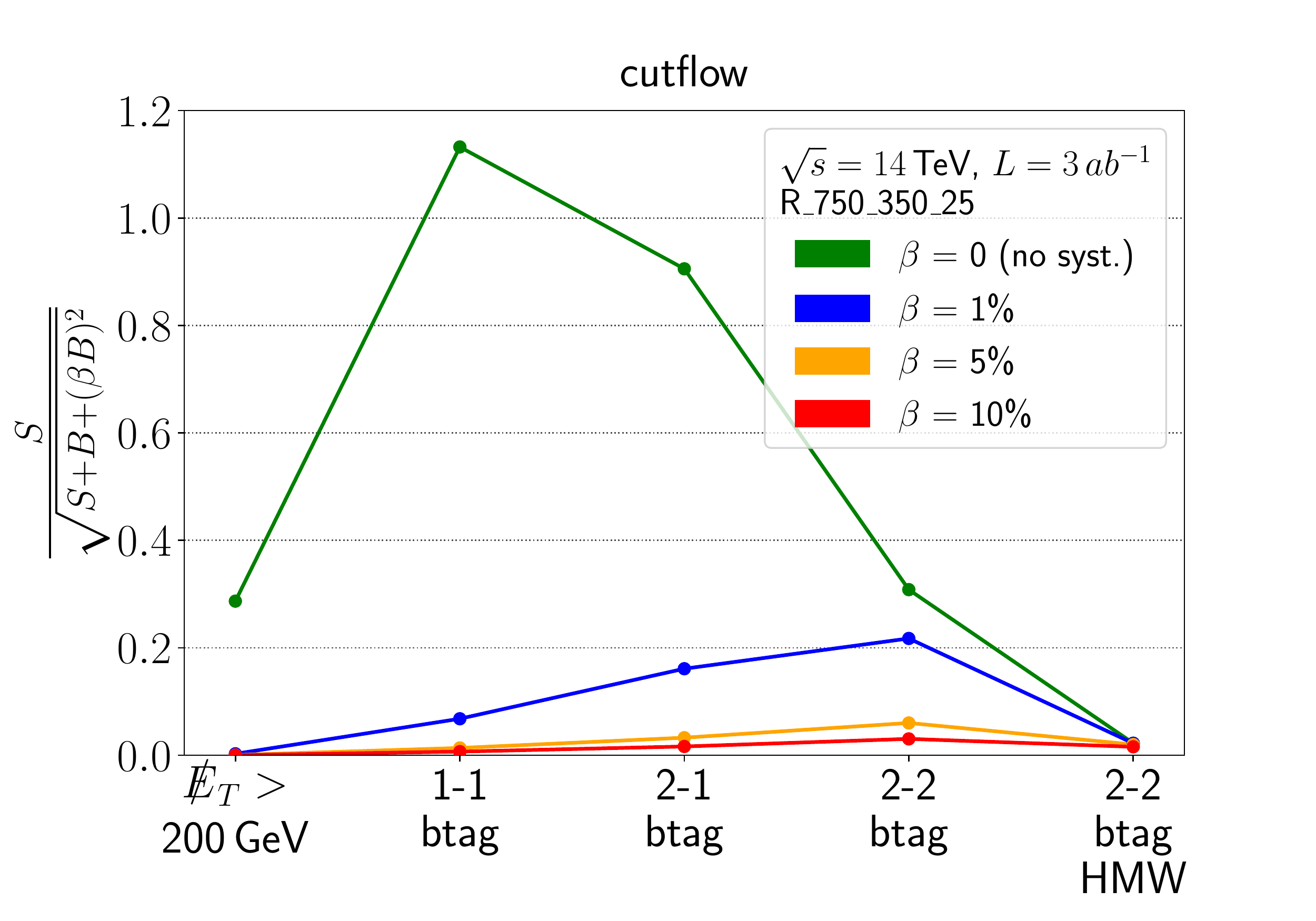} 
\end{minipage} \hfill
\begin{minipage}{.26\textwidth}
\vspace*{-0.30cm}
 \begin{tabular}{|l|c|}
 \hline
  {\bf backgd. ($R=0.6$)} & {\bf\boldmath $\sigma$ [fb]} \\\hline
  \hline
  $Z(\rightarrow\nu\bar{\nu})\,b\bar{b}jj$ & 356 \\\hline
  $W(\rightarrow\ell\nu)\,jjjj$ & 164 \\\hline
  $W(\rightarrow\ell\nu)\,b\bar{b}jj$ & 158 \\\hline
  $t\,(\to bjj)\,\bar{t}\,(\to \bar{b}\ell\nu)$  & 126\\\hline
  $Z(\rightarrow\nu\bar{\nu})\,jjjj$ & 113 \\\hline
  $Z(\rightarrow\nu\bar{\nu})\,b\bar{b}b\bar{b}$ & 4 \\
\hline
 \end{tabular}
\end{minipage}
\caption{\label{fig:cutflowReso}Cutflow (left panel) and significance $\Sigma$ (center panel) for the cut-based analysis of the resonant model $R\textunderscore750\textunderscore350\textunderscore25$. Right panel: Cross section of dominant backgrounds for $\met > 200\gev$, after applying a lepton veto and requiring 1-1 $b$-tagging, i.e., at least one $b$-tagged subjet from each fat jet with radius $R=0.6$.}
\end{figure}

The cutflow analysis is summarized for the symmetric benchmark $S\textunderscore750\textunderscore350\textunderscore25$ in Figure~\ref{fig:cutflowChain} and for the resonant benchmark $R\textunderscore750\textunderscore350\textunderscore25$ in Figure~\ref{fig:cutflowReso}. In the right panel of each figure, we list the cross section for the dominant background processes for $\met > 200\gev$ and after applying a lepton veto and requiring at least two $b$-tagged subjets from fat jets with radius $R=1.2$ (symmetric topology) and $R=0.6$ (resonant topology), respectively. We have checked that contributions from di-boson plus jets production are smaller. The latter will be neglected in our analysis.
%

To discriminate between our signal and the backgrounds, we apply a cut-and-count procedure. Throughout our analysis, we apply an initial cut of $\met > 200\gev$ and a lepton veto. To study the impact of our $b$-tagging technique, we request various $x$-$y$ $b$-tags with $0 \le x,y \le 2$, one by one.\footnote{The upper limit on $x,y$ is due to the fact that we only consider the two hardest subjets within a given fat-jet.} We furthermore apply an optional Higgs Mass Window (HMW) by requesting that the mass of each of the two identified fat jets lies within the window $75\gev < m_J < 175\gev$. In addition, we allow for a variable lower cut on $\met$ (in steps of 50 GeV).\footnote{For the symmetric topology, a looser $\met$ cut is preferred to optimize the significance. For the resonant topology, due to the harder $\met$ spectrum and the low number of signal events, the preferred cut lies at higher $\met$. In order to establish a fair comparison of the remaining selection criteria in the two topologies, we discuss the cutflow for a fixed cut of $\met > 200\gev$.} To optimize the choice of the jet radius, we have carried out our analysis for four different fat-jet radii $R=0.6,\,0.8,\,1.0,\,1.2$. For the symmetric topology, the significance is maximized for $R=1.2$, while the resonant model favors a smaller fat-jet radius of $R=0.6$.\footnote{For the resonant model, the sensitivity with $R=0.6$ is a factor of 2 higher than for $R=1.2$.}

The impact of the various cuts on signal and background is shown for the symmetric model in the left panel of Figure~\ref{fig:cutflowChain}. We see that applying $2$-$2$ $b$-tags plus a Higgs mass window leaves us with about 40 signal events, while the sum of backgrounds ranges around 700 events. In the cut-and-count analysis, we therefore do not achieve a signal-to-background ratio of ${\cal O}(1)$, so that the significance depends critically on systematic uncertainties that can affect the analysis. In the center panel, we illustrate this dependence by showing the significance defined as
\begin{align}\label{eq:significance}
\Sigma = \frac{S}{\sqrt{S + B + (\beta B)^2}}.
\end{align}
We assume a systematic uncertainty of $\beta = 0, 1, 5, 10 \%$, respectively. For a small uncertainty $\beta = 1\%$, a maximum significance of about $\Sigma = 2$ can be reached for the considered benchmark. In the resonant model, shown in Figure~\ref{fig:cutflowReso}, the signal rate is significantly lower than in the symmetric model. With our basic cut-and-count analysis, we therefore do not achieve a noticeable sensitivity to our signal. The fact that the significance of our signal depends on a combination of various kinematic variables suggests to perform a multi-variate analysis to optimize the sensitivity.

%% file: mva.tex
\noindent In this section, we describe the strategy pursued in our multi-variate analysis (MVA) and present our results. Depending on the respective phase-space point, discriminating between the di-Higgs plus $\met$ signal and the backgrounds can be  
very challenging. In order to maximize the sensitivity for each of our
benchmarks in the two models, we perform a multi-variate analysis. In this study, we use the 
{\tt scikit-learn}~\cite{Pedregosa:2012toh} implementation of {\tt AdaBoost}~\cite{Freund:1997xna}, employing the {\tt SAMME.R} algorithm to perform a Boosted Decision Tree (BDT) classification. As our best setup, we choose to train 70 trees with a maximal depth of 3, a learning rate of 0.5 and a minimum node size of 0.025 of the total weights.

Before running the BDT, we place basic kinematic selection cuts on the missing transverse energy and the jets. As in the cutflow analysis, we apply a $\met > 200\gev$  cut and veto events containing isolated leptons.\footnote{We require electrons (muons) to have $p_{T\text{lep}}>10\,(10)$~GeV 
and $|\eta_\text{lep}|<2.5 (2.7)$, and consider them isolated if $\frac{p_{T\text{lep}}}{p_{T\text{had}}}<0.12\,(0.25)$ 
within $R=0.5\,(0.5)$ of the electron (muon) momentum. Looser lepton
selection criteria might result in a better rejection of the $W$+jets background 
and therefore improve the significance of our analysis. However, in this work we consistently use our
conservative lepton definition.} The jets are defined as Cambridge-Aachen fat jets $J$ with a jet radius of $R=1.2$ ($R=0.6$) for the symmetric (resonant) topology and transverse momentum $p_T(J) > 20 \gev$. A fat jet $J_i$ is accepted if it contains two subjets $j_i^k$, $i,k=\{1,2\}$, where at least one of them is $b$-tagged.

In our multi-variate analysis, we use kinematic information on the two hardest $b$-tagged fat jets, $J_1$ and $J_2$ and their corresponding subjets $j_1^k$ and $j_2^k$. The complete set of variables used for our analysis can be classified in four categories:
\begin{itemize}
\item Global variables: missing transverse energy, $\met$; $H_T$ (computed using the fat jets); total number of fat jets, $N_J$; total number of $b$-tagged fat jets, $N_{Jb}$; total number of $b$-tagged subjets within all fat jets, $N_{jb}$;
\item Single fat-jet variables: transverse momentum $p_T(J_i)$; pseudo-rapidity $\eta(J_i)$; jet mass $m_{J_i}$; azimuthal angular separation between fat jet and missing momenta, $\Delta \phi (J_i, \met)$; ratio of transverse momenta $p_T(J_i) / \met$;
\item Two fat-jet variables: distance between two fat jets, $\Delta R(J_1,J_2)$; invariant mass of two fat jets, $m_{J_1 J_2}$; maximum jet mass ratio, max($m_{J_1} / m_{J_2},m_{J_2} / m_{J_1}$);
\item Subjet variables: transverse momentum $p_T (j^{k}_i)$; pseudo-rapidity $\eta (j^{k}_i)$; distance between subjets, $\Delta R(j^{k}_i,j^{l}_i)$.
\end{itemize}
We employ $80\%$ of our events for training and $20\%$ for evaluation purposes. The different backgrounds are weighted according to their relative cross section after applying the basic selection cuts. The BDT thus focuses on the dominant backgrounds when trained to avoid misidentification of the respective backgrounds as a signal. To make sure that the BDT will take equal effort in correctly classifying the overall number of signal events and background events, we scale the total weight of all signal events to match the total weight of all background events. This is especially important, since in the training we involve more 
Monte Carlo events for the background processes than for the signal. The BDT will assign a \emph{score} (or threshold in machine learning (ML) terminology) to each event, which reflects the likelihood of it being signal.
 
\begin{figure}[t]
\centering
\includegraphics[height=2.5in]{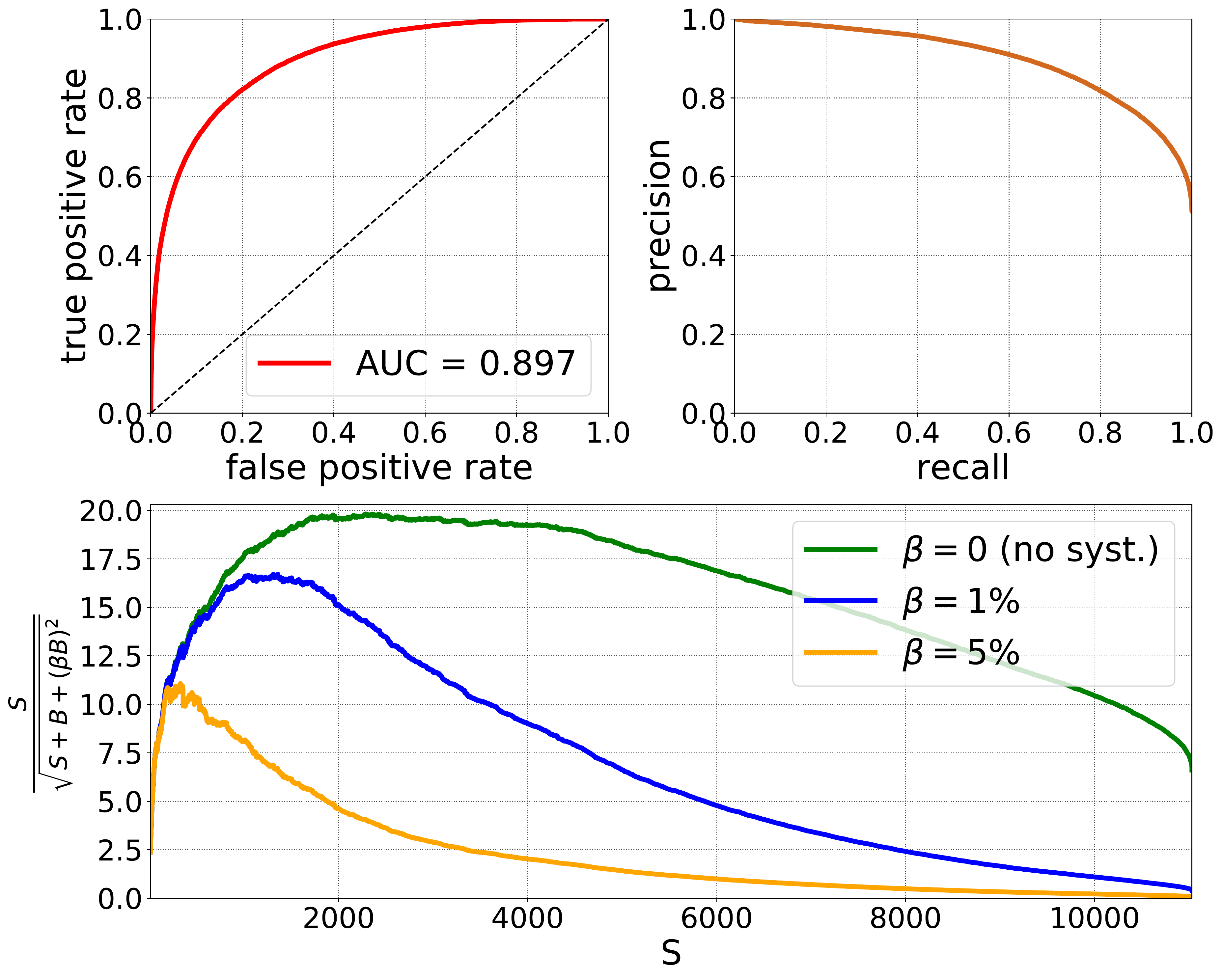} \hspace*{0.5cm}
\includegraphics[height=2.5in]{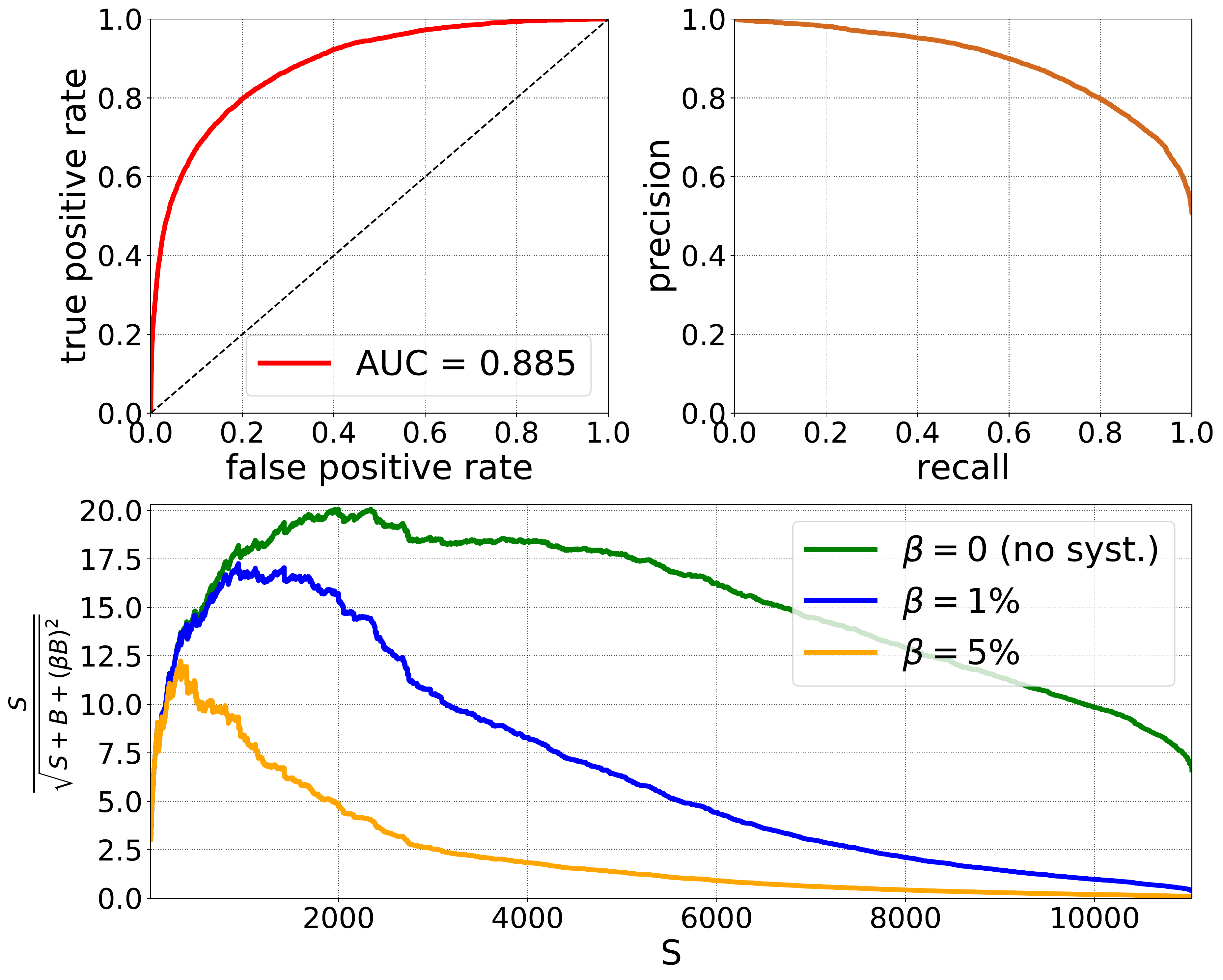} \hspace*{0.5cm}
\caption{\label{fig:rocS}Training (left) and evaluation (right) results of the BDT analysis for the symmetric benchmark $S\textunderscore750\textunderscore350\textunderscore25$, using a fat-jet radius of $R=1.2$. The lower panels show the significance $\Sigma$ as a function of the expected signal events $S$ for $\sqrt{s} = 14\tev$ and $L=3\,\text{ab}^{-1}$.}
\end{figure}

In Figure~\ref{fig:rocS}, we show a typical BDT result for the symmetric benchmark $S\textunderscore750\textunderscore350\textunderscore25$.
 The three plots to the left show the training results, while the three plots to the right display the outcome of the evaluation. In the upper left plot of each panel, we present the Receiver Operating Characteristic (ROC) curve. Shown is the \emph{true positive rate} (TPR, also referred to as \emph{recall} in ML language) against the \emph{false positive rate} (FPR, also referred to as fall-out).~\footnote{The TPR is the probability that a signal event gets tagged as signal, while the FPR is the probability that a background event gets tagged as signal.} The most relevant information is the Area Under the Curve (AUC), which quantifies the BDT capability to discriminate between signal and background. By construction, the AUC ranges between 0.5 and 1. For all our benchmarks, the AUC is at or above $0.9$, which proves that our signal/background classifier has an impressive performance. 
 
The upper right plot of each panel displays the \emph{precision} (or positive predictive value) as a function of the recall. The precision is defined as the fraction of true signal events among those events the BDT classified as signal. 
This curve illustrates how reliable a classification as signal is, depending on which fraction of signal events are classified correctly. In the displayed curve, we used the adjusted signal weights, ensuring that signal and background are on equal footing (see the discussion above). Hence the minimum value for the precision is 0.5. 

The lower plot in each panel shows the significance $\Sigma$ defined in Eq.~(\ref{eq:significance}) as a function of the expected number of signal events $S$ that would be left after cutting on a given score. The prediction is made for the HL-LHC with an integrated luminosity of $L=3\,\text{ab}^{-1}$. We present the significance for this benchmark for three different assumptions of systematic uncertainties. It is apparent that the use of a BDT enhances the significance by about an order of magnitude compared to our basic cut-and-count analysis.\footnote{\label{fn:spikes}The small spikes in the significance curve in the evaluation sample are due to a lack of Monte Carlo statistics in the $Wjjjj$ background for high BDT scores. We have simulated $10^8$ Monte Carlo events for this background. Owing to the large number of colored final states, however, the event generation is computationally intense. Moreover, the lepton veto cannot be reliably implemented at parton level, which enhances the number of required events. Generating a larger sample of $Wjjjj$ events would soften the spikes, but is beyond the scope of our analysis.}

In order to determine the sensitivity to a given benchmark scenario, we cut the BDT score (in our case, the number of signal events) at the peak of the evaluation significance. In our example $S\textunderscore 750\textunderscore 350\textunderscore 25$ and assuming $5\%$ systematics, this corresponds to a significance of $\Sigma = 12$ and $S = 325$ and $B = 238$ remaining signal and background events, compared to $S=42$ and $B=713$ in the basic cut-and-count approach. The  increase in sensitivity is due to a better selection of signal events, while at the same time having a similar improvement in background rejection. All of our benchmarks in the symmetric model feature a signal-to-background ratio close to unity, which suggests good discovery prospects in the presence of systematic uncertainties. To take into account the statistical uncertainty on our $Wjjjj$ background simulation (see footnote~\ref{fn:spikes}), in what follows we take a conservative approach and claim a ``discovery'' at a significance of $\Sigma=7$ (corresponding to 7 standard deviations from the standard-model hypothesis for Gaussian statistics), instead of the common $5\sigma$ threshold.

\begin{figure}[t]
\centering
\includegraphics[height=2.5in]{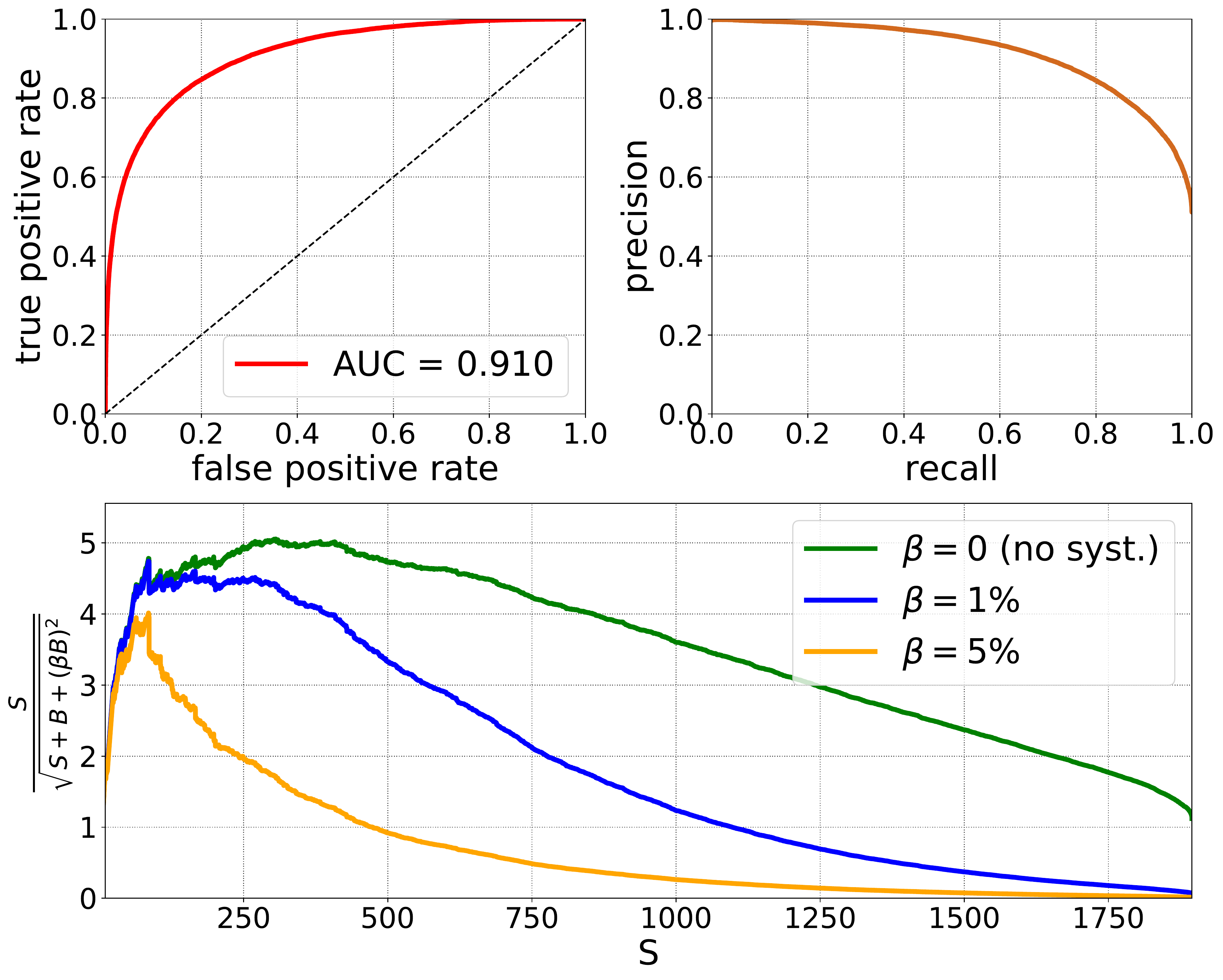} \hspace*{0.5cm}
\includegraphics[height=2.5in]{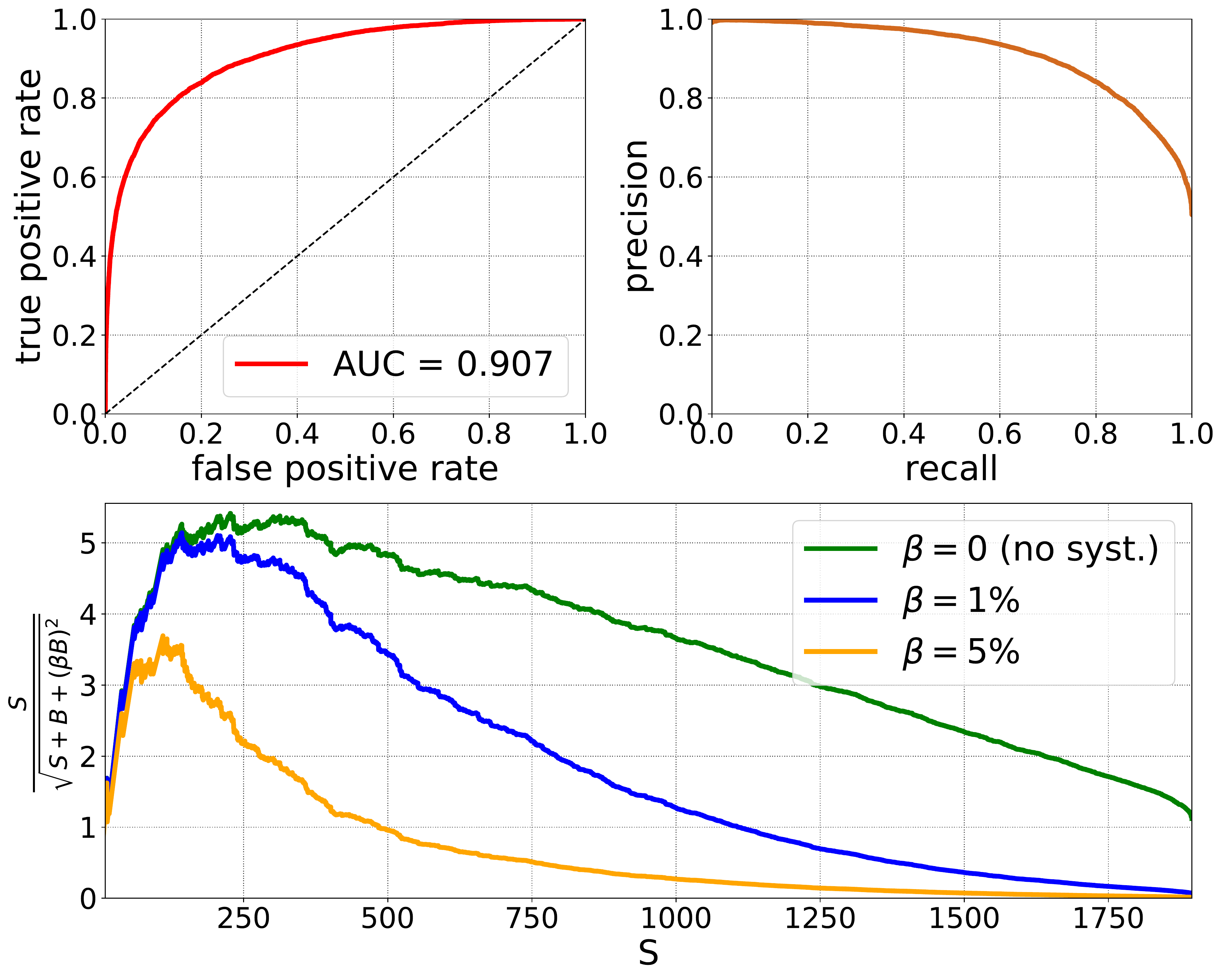} \hspace*{0.5cm}
\caption{\label{fig:rocR}Training (left) and evaluation (right) outcome of our BDT for the resonant benchmark $R\textunderscore750\textunderscore350\textunderscore25$, using a fat-jet radius of $R=0.6$. The lower panels show the significance $\Sigma$ as a function of the expected signal events $S$ for $\sqrt{s} = 14\tev$ and $L=3\,\text{ab}^{-1}$.}
\end{figure}

In Figure~\ref{fig:rocR}, we present our results for the resonant benchmark $R\textunderscore750\textunderscore350\textunderscore25$.
 Again, we see the excellent performance of our BDT classifier. However, in the resonant model the lower signal rate severely limits the sensitivity. While the BDT improves the sensitivity, we cannot reach the discovery level for our models. Still, a significance of $\Sigma \approx 2$ can be achieved in benchmarks with a lighter scalar $B$, corresponding to a larger signal rate. The HL-LHC can thus test parts of the parameter space, but a more refined strategy (or a combination of multiple channels) would be required to reach a higher sensitivity. In summary, compared with the cut-and-count analysis the BDT enhances the sensitivity to the resonant topology by about a factor of 10 in most benchmarks. Still, the sensitivity is lower than for the symmetric topology, mostly due to the reduced signal rate.

\begin{figure}[t]
\centering
\includegraphics[width=0.46\textwidth]{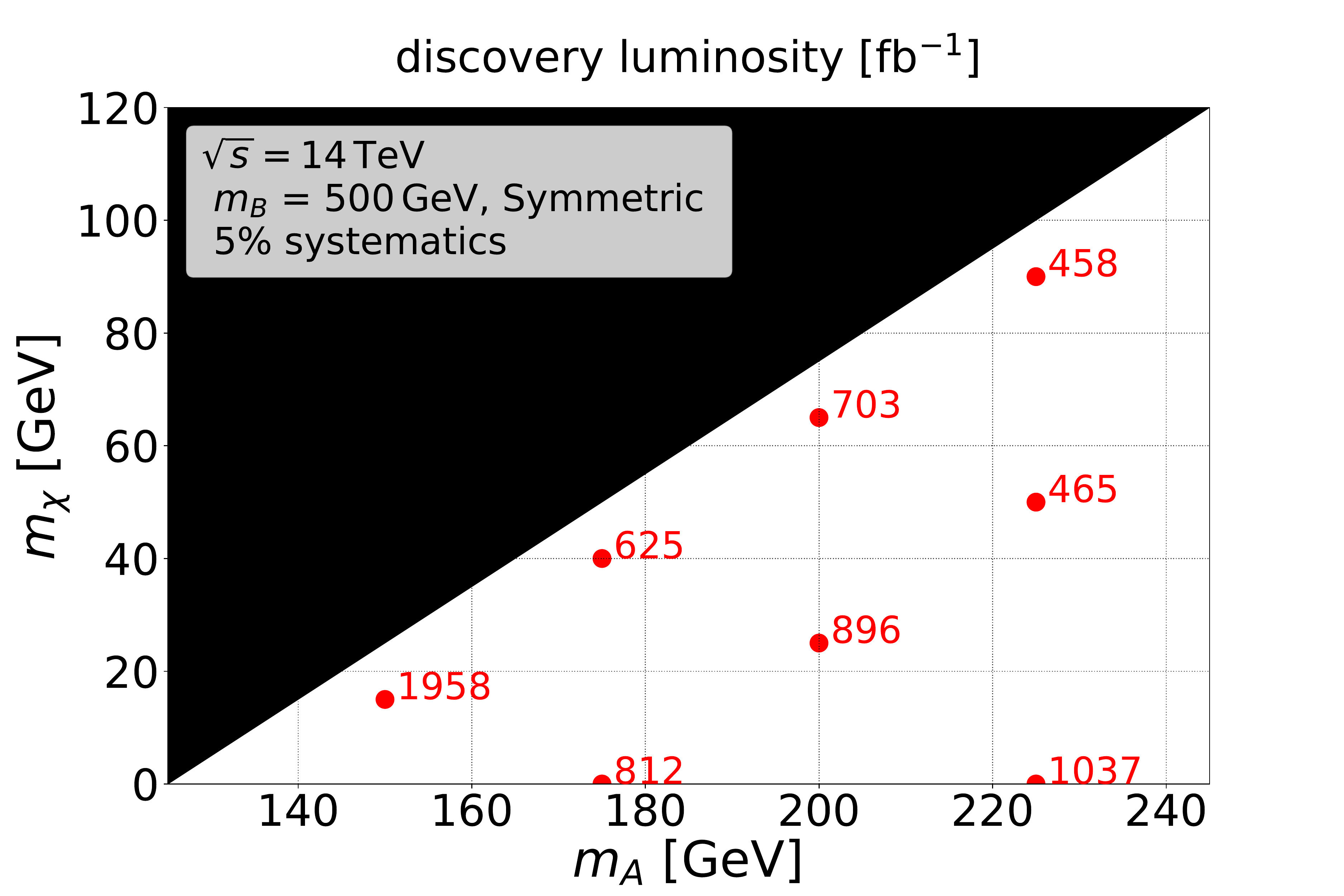} \hspace*{0.5cm}
\includegraphics[width=0.46\textwidth]{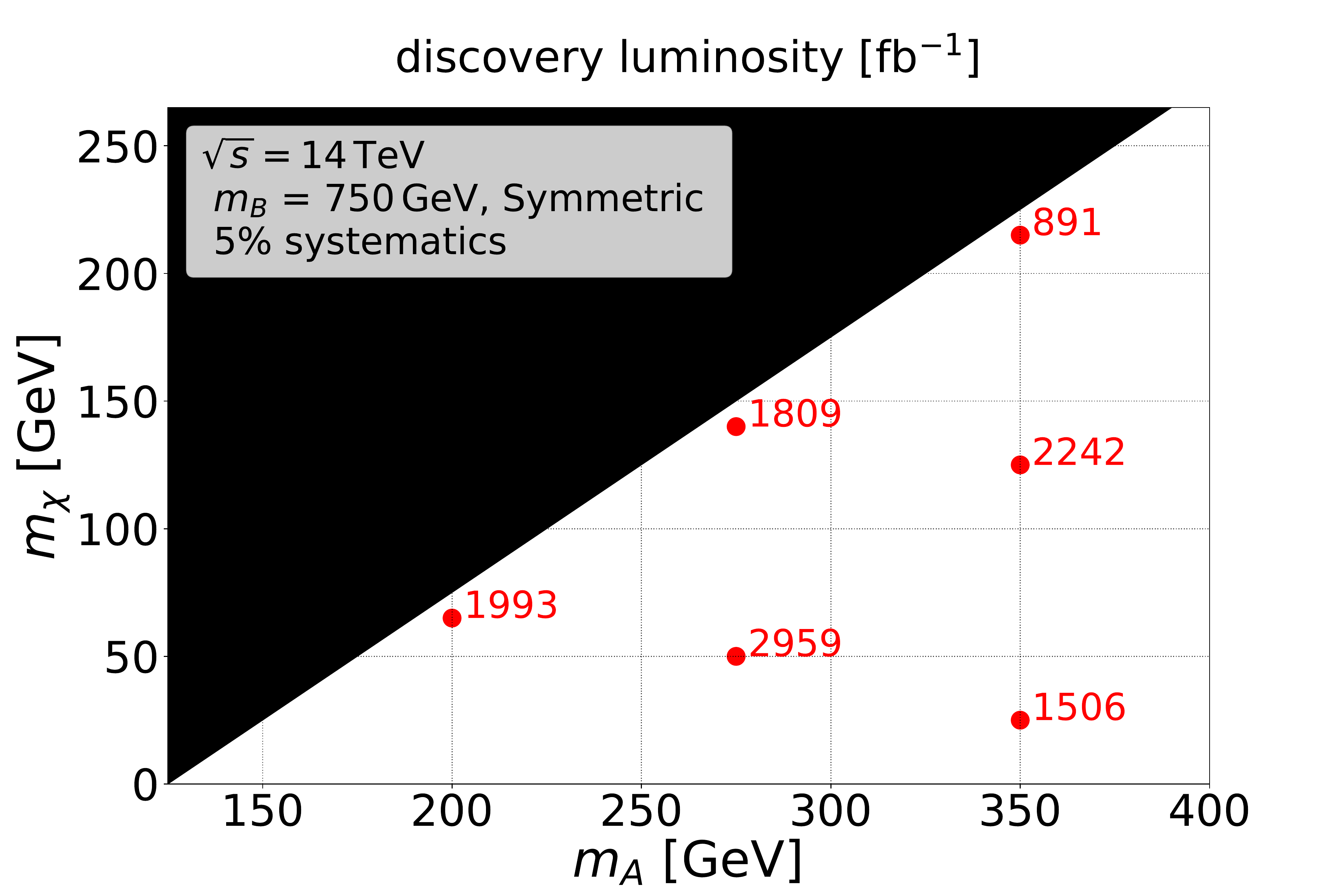} \hspace*{0.5cm}
\caption{\label{fig:discolumiS}Luminosity required for a discovery (in fb$^{-1}$) at the HL-LHC in the $m_A - m_\chi$ plane for the symmetric model, with $m_B = 500\gev$ (left panel) and $m_B = 750\gev$ (right panel).}
\end{figure}

To show the dependence of the signal sensitivity on the respective model parameters, we present our results in terms of the scalar masses $m_A$ and $m_\chi$. In Figure~\ref{fig:discolumiS}, we display the luminosity required to discover the symmetric benchmark scenarios from Table~\ref{tab:benchChain} at the HL-LHC, assuming a systematic uncertainty of $5 \%$. The mass of the heavy scalar is fixed to $m_B=500\gev$ (left panel) and $m_B = 750\gev$ (right panel).
Apart from one benchmark, all scenarios are well within the reach of the HL-LHC. We also see that the significance is particularly high for benchmarks with a compressed spectrum. As anticipated from Figure~\ref{fig:met_distro_topo}, for $m_B \gtrsim 2m_A$, the $\met$ spectrum is harder and the cut on missing energy is thus more efficient in rejecting the background. In Figure~\ref{fig:discolumiS}, this effect can be seen by looking at fixed values of $\Delta_{Ah\chi}$ and increasing $m_A$ (along the diagonal). For a more compressed spectrum, the required luminosity is drastically reduced by a factor of 10-20, depending on the actual value of $m_B$. A similar but milder effect occurs if the decay $A\to h\chi$ proceeds close to threshold. In this case, the $h\chi$ pair follows the direction of $A$, resulting again in a harder $\met$ distribution. In the plot, this corresponds to fixed values of $m_A$ and increasing $m_\chi$ (along the vertical direction), resulting in a reduction of the required luminosity by a factor of about 2-3 at most.  In summary, a di-Higgs signal could be discovered in an early phase of the HL-LHC, provided that the scalar resonance $B$ is produced at a sizeable rate.

\begin{figure}[t]
\centering
\includegraphics[width=0.46\textwidth]{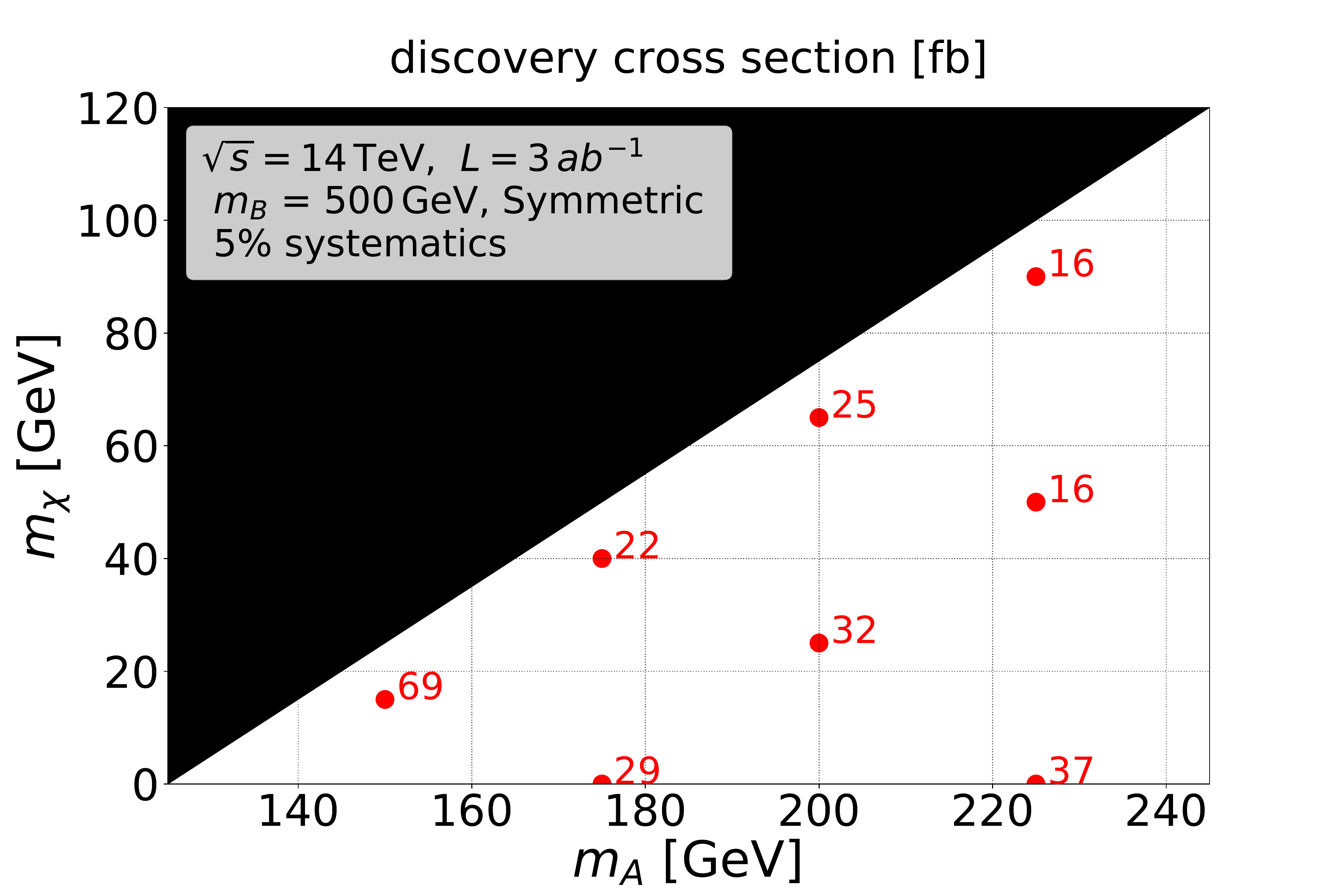} \hspace*{0.5cm}
\includegraphics[width=0.46\textwidth]{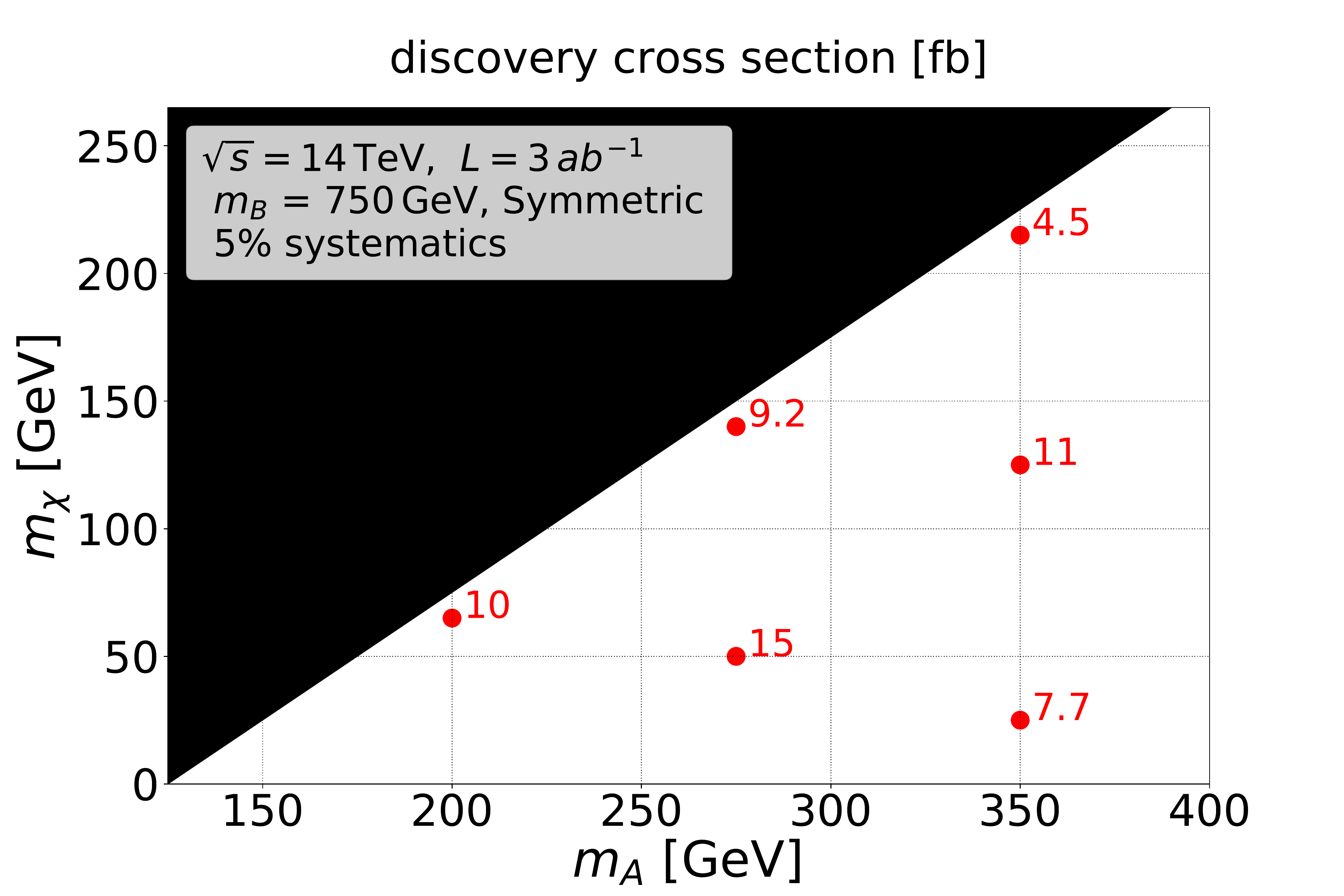} \hspace*{0.5cm}
\caption{\label{fig:S750xs}Cross sections (in fb) required for a discovery at the HL-LHC in the $m_A - m_\chi$ plane for the symmetric model, with $m_B = 500\gev$ (left panel) and $m_B = 750\gev$ (right panel).}
\end{figure}

We present our results in a second way, which is particularly convenient for recasting purposes. In Figure~\ref{fig:S750xs}, we report the cross section that can be probed at the discovery level for a luminosity of 3 ab$^{-1}$. Again, we fix the heavy scalar mass at $m_B= 500\gev$ (left panel) and $m_B= 750\gev$ (right panel), respectively.
We see that in the most difficult benchmark topology, namely in the benchmark that requires the largest discovery luminosity at the HL-LHC,  $S\textunderscore 750\textunderscore 275\textunderscore 50$, a cross section of about 15~fb would be required to claim a discovery. It is interesting to compare this value with the latest di-Higgs predictions for the SM that require a total rate for $hh \to 4b$ of about 13~fb for discovery. In the standard model, the Higgs pair is not produced through a resonance, and furthermore the final state of four $b$-quarks without $\met$ is difficult to identify. In contrast, in our scenarios the scalar $B$ is resonantly produced and the large $\met$ in the final state facilitates an efficient discrimination against the backgrounds. The fact that we can probe cross sections of a few to several femtobarns is an important result, which motivates a study of the di-Higgs plus $\met$ signature in the context of complete models. For the benchmarks with $m_B = 1\tev$, the signal rates are too low to claim a discovery at the HL-LHC with $3$~ab$^{-1}$. However, already with slightly more luminosity (or larger cross section) a discovery of these heavy scalar scenarios is possible.

\begin{figure}[t]
\centering
\includegraphics[width=0.46\textwidth]{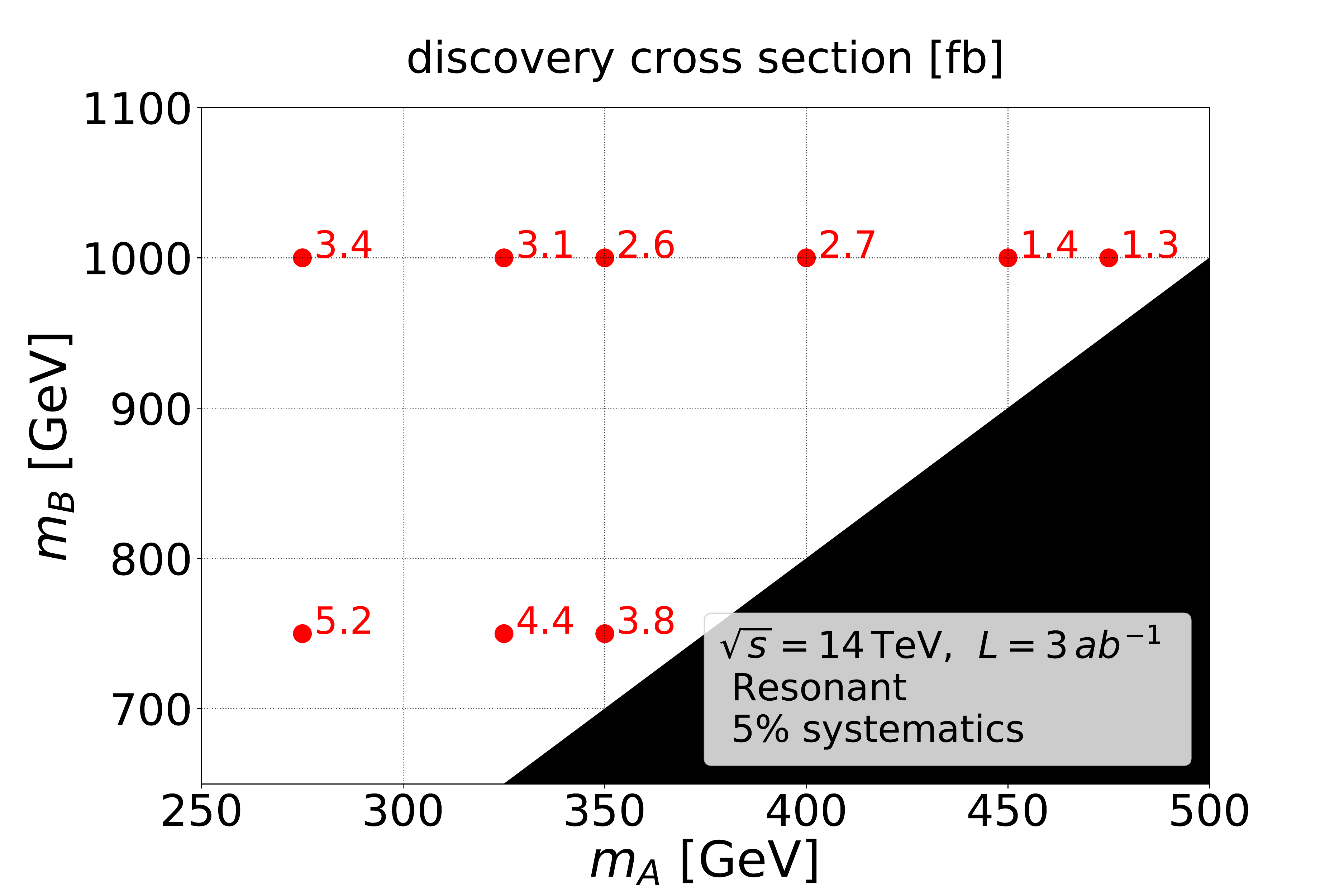} \hspace*{0.5cm}
\includegraphics[width=0.46\textwidth]{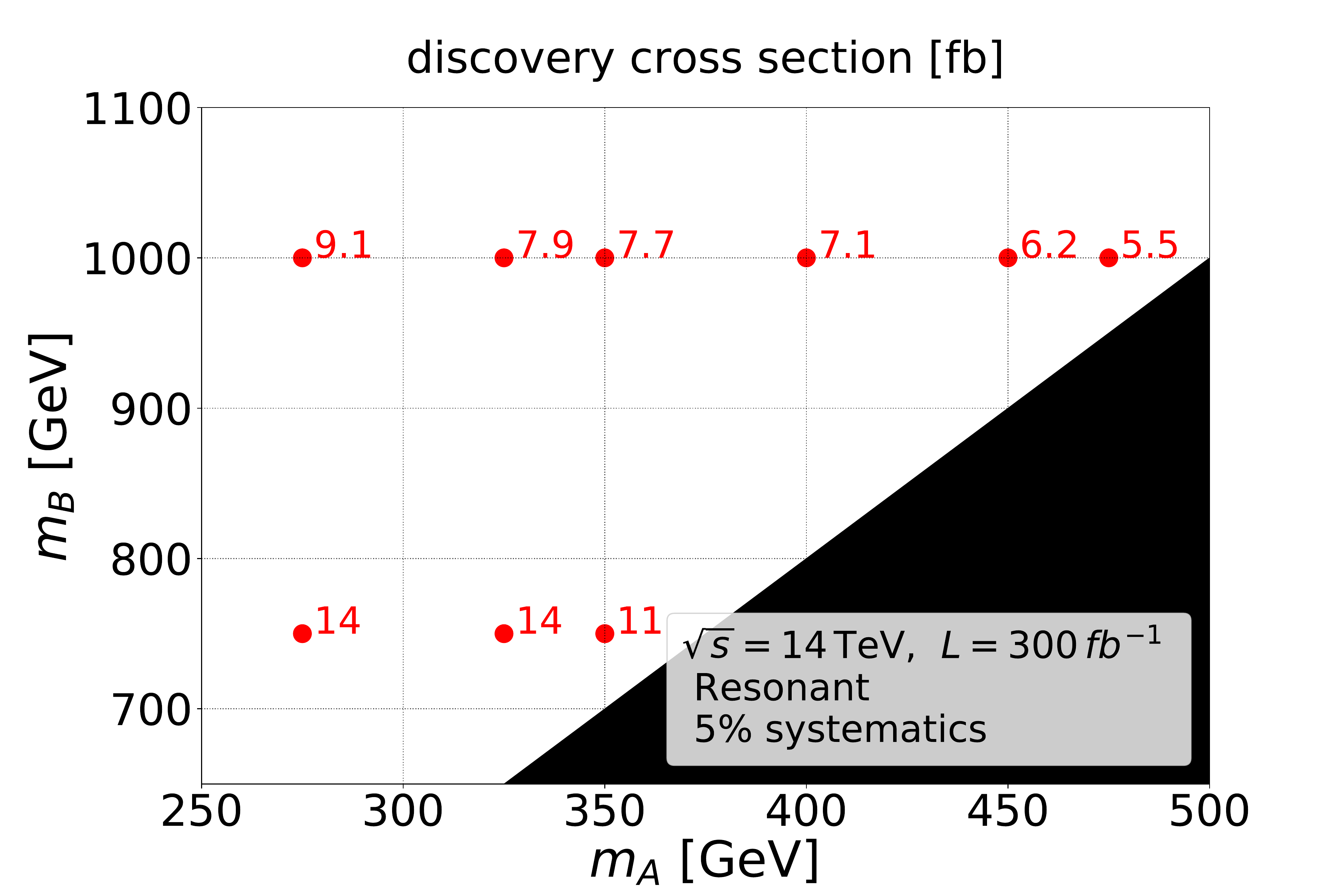} \hspace*{0.5cm}
\caption{\label{fig:Rxs}Cross sections (in fb) required for a discovery at the HL-LHC in the $m_A - m_\chi$ plane for the resonant model, with $3\,\text{ab}^{-1}$ (left panel) and $300\,\text{fb}^{-1}$ (right panel). We have fixed $m_\chi =25\gev$, but as explained in the text this parameter is not relevant for the sensitivity provided that $2m_\chi < m_A$.}
\end{figure}

In the resonant model, the planned HL-LHC luminosity is not sufficient to discover any of the benchmark scenarios, due to the lower production rates. We therefore confine ourselves to presenting the cross sections required to discover a particular resonant benchmark in Figure~\ref{fig:Rxs}. As explained in Section~\ref{sec:signature}, the mass of the lightest scalar, $m_\chi$, does not affect the sensitivity, since the boost of $A$ does not depend on $m_\chi$ or $m_h$. We therefore present our results in terms of the heavier scalar masses $m_A$ and $m_B$.
 From the figure, we see that we can only test cross sections in the femtobarn and sub-femtobarn regime. As in the symmetric model, the significance increases when the spectrum is compressed. Increasing $m_A$ for fixed $m_B = 1000\gev$ lowers the testable cross section by a factor of $3$. Using {\tt CheckMATE2} we have verified that even with the largest possible cross section displayed here, the search for di-Higgs plus $\met$ is still the most sensitive channel for the resonant model.

%% file: dm.tex
\noindent In our models, the lightest scalar $\chi$ is automatically stable, due to the imposed $\mathbb{Z}_2$ symmetry. Here we explore the hypothesis that $\chi$ is a dark matter candidate. We discuss the dark matter phenomenology for our simplified models, focusing on dark matter-nucleon scattering in direct detection experiments and thermal freeze-out in the early universe.

A contribution to spin-dependent nucleon scattering arises from the portal operator $H^\dagger H\,\chi\chi$, which induces a $h\chi\chi$ interaction after electroweak symmetry breaking. The effect of this operator has been well studied elsewhere (see e.g. Ref.~\cite{Athron:2017kgt} for a recent study). Since it is not relevant for our di-Higgs plus $\met$ final state, we will assume that it is absent. We follow the same philosophy for other operators, namely we only study the implications of operators that play a role in the collider phenomenology.

 In the symmetric model, $A$ and $\chi$ can mix through the coupling $\lambda_{A\chi HH}$ from Eq.~(\ref{eq:int-s}) upon electroweak symmetry breaking. This mixing induces a Higgs coupling to the lightest scalar, $\lambda_{A\chi HH}\sin\theta_{A\chi}$, where $\theta_{A\chi}$ is the $A$-$\chi$ mixing angle. Direct detection experiments set a strong bound on this coupling. However, as we explained in Section~\ref{sec:models}, neither $\lambda_{A\chi HH}$ nor the mixing affects the signal strength of di-Higgs plus $\met$. We have therefore set $A$-$\chi$ mixing to zero in our analysis, $\theta_{A\chi} = 0$. The lightest scalar $\chi$ can be a viable dark matter candidate that leaves a di-Higgs plus $\met$ signature at the LHC while agreeing with the null results from direct detection experiments.

In the resonant model, $A$ and $h$ mix through the operator $m_{AHH} A H^{\dagger} H$ after electroweak symmetry breaking. For fixed $m_A$ and $m_\chi$, direct detection experiments set an upper bound on the product of the $m_{A \chi \chi}$ and $m_{A H H}$ couplings from mixing-induced nucleon scattering. Since the signal strength of di-Higgs plus $\met$ production depends on the relative size of $m_{AHH}$ and $m_{A\chi\chi}$, but not on their overall magnitude (see Section~\ref{sec:models}), we conclude that in the resonant model the lightest scalar $\chi$ is not ruled out as a dark matter candidate by direct detection experiments if $m_{A \chi \chi} m_{A H H}$ is sufficiently small.

Assuming that our dark matter candidate is a relic from thermal freeze-out in the early universe sets additional constraints on the parameter space of our models. In the symmetric model, dark matter annihilation can be efficient in either of the following scenarios,\footnote{Additional annihilation processes such as  $\chi \chi \to B \to  gg$ via the $C_{Bgg}$ coefficient are possible. However, such channels would also require a $B\chi\chi$ interaction, which is irrelevant for the di-Higgs plus $\met$ signature. As in the case of direct detection, we will neglect these interactions.}
\begin{align}
m_\chi > m_h:\quad & \chi\chi \to hh\quad\qquad \sim \lambda_{A\chi HH}^4,\\\nonumber
2m_\chi \approx m_h:\quad & \chi\chi \to h \to b\bar{b}\quad\, \sim\lambda_{A\chi HH}^2 \sin^2\theta_{A\chi}.
\end{align}
If dark matter is heavier than the Higgs boson, it can annihilate by $t$-channel $A$ exchange, scaling as $\lambda_{A\chi HH}^4$. The observed dark matter abundance of $\Omega_\chi h^2 = 0.1199$~\cite{Ade:2015xua} can be obtained with moderate couplings and mediator masses $m_A \lesssim 1\tev$. If dark matter is lighter than the Higgs boson, it can only annihilate through $A$-$\chi$ mixing, which is strongly suppressed by the null results from direct detection experiments. It is still possible to satisfy the observed relic abundance for dark matter pair masses near the Higgs mass. In this case, $s$-channel annihilation through the Higgs boson occurs resonantly, which compensates for the coupling suppression.

In the resonant model, the direct detection bounds on $m_{A\chi\chi}\sin\theta_{Ah}$ suppress all interactions of dark matter. The observed relic abundance can only be obtained for dark matter pairs in the Higgs resonance region,
\begin{align}
2m_\chi \approx m_h:\quad & \chi\chi \to h \to b\bar{b}\quad \sim m_{A\chi\chi}^2 \sin^2\theta_{Ah}.
\end{align}
Obtaining the observed relic abundance away from the Higgs resonance requires additional dark matter annihilation channels beyond what is predicted in our simplified model. In any case, the dark matter hypothesis should not constrain the search strategy for di-Higgs plus $\met$ at the LHC. For instance, a di-Higgs plus $\met$ signature could also arise in models with hidden sectors, where $\chi$ decays visibly at a later time and outside the detector, so that its decay products could be caught by dedicated detectors such as FASER~\cite{Feng:2017uoz,Ariga:2018uku} or MATHUSLA~\cite{Evans:2017lvd,Curtin:2018mvb,Alpigiani:2018fgd}.

%% file: conclusions.tex
\noindent In this work we have developed a search strategy for $hh\chi\chi$ production at the LHC using the final state with four $b$-jets and missing transverse energy. For the purpose of our and future studies, we have built two simplified models that give rise to this final state in different kinematic topologies. Both models feature a hidden sector of three new scalar singlets $A$, $B$ and $\chi$, the latter being stable due to an imposed parity symmetry. Since the scalars $A$ are pair-produced via a resonant scalar $B$, event rates in both models are sizeable at the HL-LHC. The decay of $A$ depends on the properties of the particles in the hidden sector and determines the event topology. In our \emph{symmetric} model, both $A$ and $\chi$ belong to the dark sector, so that $A \to h \chi$ decays occur on both sides of the decay chain. In the \emph{resonant} model, only $\chi$ belongs to the hidden sector and $A$ decays via $A\to hh$ or $A\to \chi\chi$ with similar branching ratios. We stress that these simplified models can be embedded in more complete models featuring an enlarged scalar sector or other particles in the hidden sector.

To demonstrate the LHC potential to discover hidden sectors with the di-Higgs plus $\met$ signature, we have performed a full-fledged numerical analysis of the multi-$b + \met$ final state, including a detailed study of SM backgrounds and detector effects. Dominant backgrounds are due to $t\bar{t}$, as well as $Wjjjj$ and $Zjjjj$ production. Employing the inclusive $\met$ trigger for event pre-selection, we have first carried out a cut-and-count analysis, followed by a multi-variate analysis based on a boosted decision tree. Both analyses rely on the use of jet-substructure techniques in a modified version of the BDRS algorithm. In our cut-and-count analysis, we employ missing energy, jet and subjet variables, as well as flavor tags to efficiently discriminate between signal and background. With this approach, we have obtained significances of at most $2 \sigma$ for the symmetric model, while in the resonant model our analysis turned out to be insensitive. 
It is thus necessary to optimize the signal-background discrimination by optimally exploiting
all kinematic features using a multivariate analysis. In particular we have shown that the use of machine-learning tools is well suited for our analysis.

In the symmetric model, our BDT analysis predicts a significance well above $5 \sigma$ for most of our scanned points, thus opening the possibility of an (early) discovery at the HL-LHC. In the resonant model, the signal rates are significantly lower (in the sub-femtobarn regime), which reduces the sensitivity. In order to claim a discovery, an increase of our benchmark cross section by a factor of 2-10 would be needed. In any case, the enhancement from our simple cut-and-count to the multivariate analysis shows that the sensitivity to the di-Higgs plus $\met$ signal relies on a variety of kinematic features in both signal and background. The BDT is thus the appropriate approach to search for such a many-body final state in an environment with large SM backgrounds. While we have focused on the $4b + \met$ channel with the largest event rates, additional final states like $b \bar{b}\,\gamma \gamma + \met$, $b \bar{b}\, W W^* +  \met$, $b \bar{b}\, \tau^+ \tau^- + \met$ can contribute significantly to enlarge the search potential of di-Higgs plus $\met$.

In the context of dark matter, it is interesting to investigate the interplay of this collider signature with searches at direct and indirect detection experiments in complete models where the relic abundance is satisfied. Within our simplified models, parts of the parameter region could provide us with a viable thermal dark matter relic. In general, potential bounds on viable dark matter models should not limit the scope in di-Higgs plus $\met$ searches at the LHC.

So far, the LHC collaborations have searched for the di-Higgs plus $\met$ signature in a specific scenario of supersymmetry with very light invisible particles in the final state. We encourage our experimental colleagues to use our simplified models and a search strategy similar to ours to fully exploit the discovery potential of di-Higgs plus $\met$. The use of machine learning techniques is crucial to achieve a high significance, in our case an enhancement by an order of magnitude over a cut-and-count analysis. Searching for di-Higgs plus $\met$ links the efforts at the dark matter frontier with those on the di-Higgs frontier, which in the last few years have seen a spectacular development in both theory and experiment. The sensitivity to this and similar signatures will greatly benefit from merging the techniques developed for Higgs pair measurements in and beyond the standard model with missing energy searches.

%% file: UV-completion.tex
\noindent In this appendix, we introduce a minimal perturbative UV completion for the effective coupling $C_{Bgg}$ in Eq.~\eqref{eq:L}. We add a heavy quark $Q$ with mass $m_Q$, odd $\mathbb{Z}_2$ parity, and vector-like weak interactions to our model. We furthermore assume that $Q$ couples to the heavy scalar $B$ via
\begin{equation}
\mathcal{L} \supset - y_{Q} \, B \bar QQ\,.
\end{equation}
For the sake of simplicity, we take $Q$ to be an $SU(2)$ singlet with hypercharge $-1/3$. Assuming that $Q$ dominantly decays via $Q\to b\chi$, its mass is constrained by sbottom searches at the LHC \cite{Aaboud:2017wqg,Sirunyan:2017kiw}, as well as by more inclusive searches for jets plus $\met$ \cite{Aaboud:2017vwy,Sirunyan:2017cwe}. 
We estimate the current bound to be at the level of $1\tev$, but leave a more thorough investigation for future work. 

Following Ref.~\cite{Falkowski:2015swt}, integrating out $Q$ generates the following effective couplings to gluons and photons,~\footnote{If we had instead introduced $Q$ as a vector-like top partner, the coupling $C_{B\gamma\gamma}$ would be larger by a factor of four.}
\begin{equation}\label{eq:CS}
\frac{C_{Bgg}}{\Lambda} = \frac{g_s^2 y_{Q}}{48\pi^2 m_Q}\,, \qquad \frac{C_{B\gamma\gamma}}{\Lambda} =  \frac{e^2 y_{Q}}{72\pi^2 m_Q}\,,
\end{equation}
where  $g_s$ is the coupling constant of QCD and $C_{Bgg}$ is defined in Eq.~\eqref{eq:L}. The effective coupling to photons is defined analogously by
\begin{equation}
\mathcal{L} \supset \frac{C_{B\gamma\gamma}}{\Lambda} B F_{\mu\nu} F^{\mu\nu}.
\end{equation}
With $m_Q = 1\,$TeV and $y_Q=1$, we find\footnote{Notice that the naive dimensional analysis (NDA) estimate for $C_{Bgg}$, assuming a loop-induced perturbative UV completion at the scale $\Lambda = 1\,$TeV, is larger by about a factor of three, $C_{Bgg} \sim 1/(16\pi^2)$.}
\beq\label{eq:CBgg-value}
C_{Bgg} = \frac{g_s^2}{48\pi^2} \simeq 2.1\cdot 10^{-3},
\eeq
and $C_{B\gamma\gamma}$ smaller by more than two orders of magnitude. Both $C_{Bgg}/\Lambda$ and $C_{B\gamma\gamma}/\Lambda$ scale as $y_Q/m_Q$, so that the effective couplings do not change when simultaneously increasing both $m_Q$ and $y_Q$. The partial decay widths mediated by these couplings are~\cite{Falkowski:2015swt}
\begin{eqnarray}
\Gamma(B\to gg) &=& \frac{C_{Bgg}^2}{m_Q^2} \frac{2 m_B^3}{\pi} = 
\frac{g_s^4 y_Q^2 m_B^3}{1152\,\pi^5 m_Q^2} 
\,,\\
\Gamma(B\to \gamma\gamma) &=& \frac{C_{B\gamma\gamma}^2}{m_Q^2}\frac{m_B^3}{4\pi} = 
 \frac{e^4 y_Q^2 m_B^3}{20736\,\pi^5 m_Q^2} \,.
\end{eqnarray}
This should be compared with the decay width into $AA$ pairs,
\begin{equation}
\Gamma(B\to AA) = \frac{\left|m_{BAA}\right|^2}{32\pi m_B}\sqrt{1-\frac{4m_A^2}{m_B^2}}\,.
\end{equation}
The production cross section $\sigma(pp\to B)$ can be estimated by making use of the results of the LHC Higgs cross-section working group~\cite{deFlorian:2016spz}, which provides the contribution of gluon-gluon fusion to the production cross section of a heavy scalar $\hat S$ with Higgs-like couplings to quarks. These numbers can then simply be rescaled to obtain the production cross section of $B$,
\beq
\sigma(pp\to B) =  \left(
\frac{m_t C_{Bgg}}{m_Q C_{\hat Sgg}}\right)^2 \sigma(pp\to \hat S) \simeq \left(\frac{y_Q}{m_Q}\frac{\sqrt{2}m_t}{y_t}\right)^2 \sigma(pp\to \hat S) = \left(\frac{y_Q v}{m_Q}\right)^2 \sigma(pp\to \hat S)\,,
\eeq
where $m_t$ and $y_t$ are the top-quark mass and Yukawa coupling.